\documentclass[twoside,twocolumn,9pt]{article}
\usepackage{extsizes}
\usepackage[super,sort&compress,comma]{natbib}
\usepackage[version=3]{mhchem}
\usepackage[left=1.5cm, right=1.5cm, top=1.785cm, bottom=2.0cm]{geometry}
\usepackage{balance}
\usepackage{widetext}
\usepackage{times,mathptmx}
\usepackage{sectsty}
\usepackage{graphicx}
\usepackage{lastpage}
\usepackage[format=plain,justification=raggedright,singlelinecheck=false,font={stretch=1.125,small,sf},labelfont=bf,labelsep=space]{caption}
\usepackage{float}
\usepackage{fancyhdr}
\usepackage{fnpos}
\usepackage[english]{babel}
\usepackage{array}
\usepackage{droidsans}
\usepackage{charter}
\usepackage[T1]{fontenc}
\usepackage[usenames,dvipsnames]{xcolor}
\usepackage{setspace}
\usepackage[compact]{titlesec}
\usepackage{srctex}
\usepackage{hyperref}

\usepackage{color}

\definecolor{cream}{RGB}{222,217,201}

\begin{document}

\pagestyle{fancy}
\thispagestyle{plain}
\fancypagestyle{plain}{

\fancyhead[C]{\includegraphics[width=18.5cm]{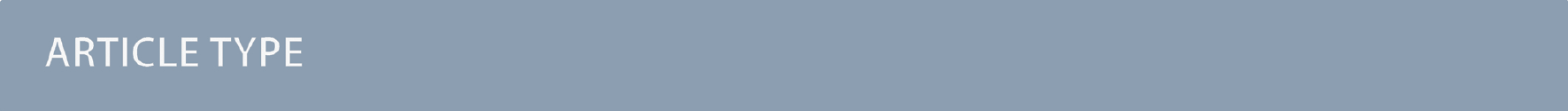}}
\fancyhead[L]{\hspace{0cm}\vspace{1.5cm}\includegraphics[height=30pt]{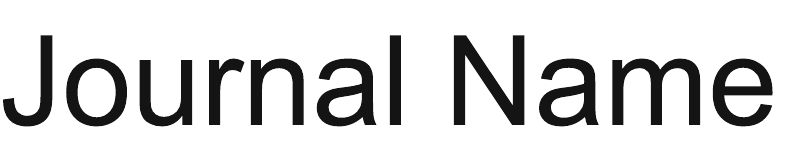}}
\fancyhead[R]{\hspace{0cm}\vspace{1.7cm}\includegraphics[height=55pt]{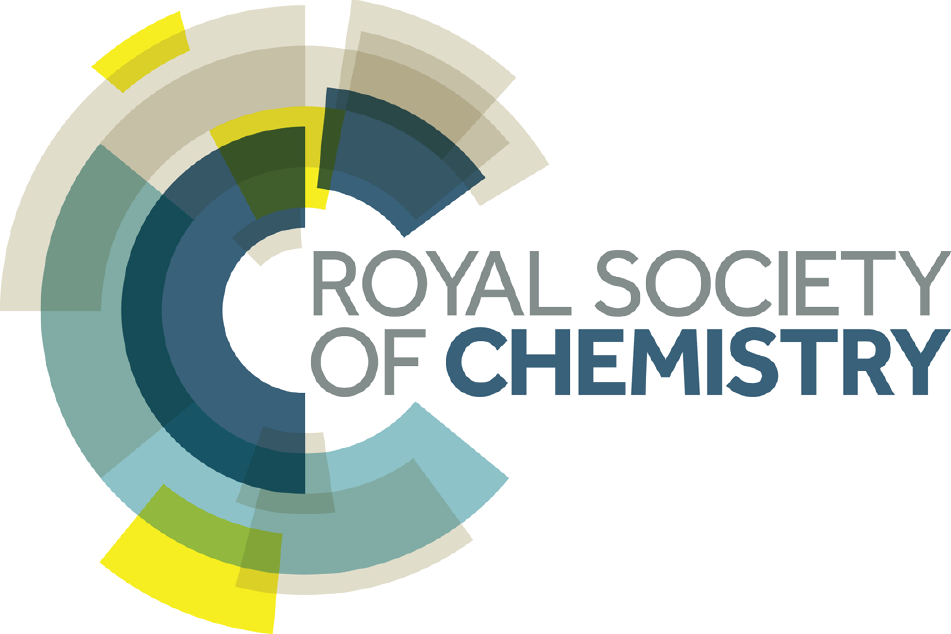}}
\renewcommand{\headrulewidth}{0pt}
}

\makeFNbottom
\makeatletter
\renewcommand\LARGE{\@setfontsize\LARGE{15pt}{17}}
\renewcommand\Large{\@setfontsize\Large{12pt}{14}}
\renewcommand\large{\@setfontsize\large{10pt}{12}}
\renewcommand\footnotesize{\@setfontsize\footnotesize{7pt}{10}}
\makeatother

\renewcommand{\thefootnote}{\fnsymbol{footnote}}
\renewcommand\footnoterule{\vspace*{1pt}%
\color{cream}\hrule width 3.5in height 0.4pt \color{black}\vspace*{5pt}}
\setcounter{secnumdepth}{5}

\makeatletter
\renewcommand\@biblabel[1]{#1}
\renewcommand\@makefntext[1]%
{\noindent\makebox[0pt][r]{\@thefnmark\,}#1}
\makeatother
\renewcommand{\figurename}{\small{Fig.}~}
\sectionfont{\sffamily\Large}
\subsectionfont{\normalsize}
\subsubsectionfont{\bf}
\setstretch{1.125} 
\setlength{\skip\footins}{0.8cm}
\setlength{\footnotesep}{0.25cm}
\setlength{\jot}{10pt}
\titlespacing*{\section}{0pt}{4pt}{4pt}
\titlespacing*{\subsection}{0pt}{15pt}{1pt}

\fancyfoot{}
\fancyfoot[LO,RE]{\vspace{-7.1pt}\includegraphics[height=9pt]{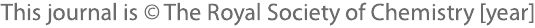}}
\fancyfoot[CO]{\vspace{-7.1pt}\hspace{13.2cm}\includegraphics{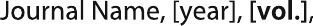}}
\fancyfoot[CE]{\vspace{-7.2pt}\hspace{-14.2cm}\includegraphics{RF}}
\fancyfoot[RO]{\footnotesize{\sffamily{1--\pageref{LastPage} ~\textbar  \hspace{2pt}\thepage}}}
\fancyfoot[LE]{\footnotesize{\sffamily{\thepage~\textbar\hspace{3.45cm} 1--\pageref{LastPage}}}}
\fancyhead{}
\renewcommand{\headrulewidth}{0pt}
\renewcommand{\footrulewidth}{0pt}
\setlength{\arrayrulewidth}{1pt}
\setlength{\columnsep}{6.5mm}
\setlength\bibsep{1pt}

\makeatletter
\newlength{\figrulesep}
\setlength{\figrulesep}{0.5\textfloatsep}

\newcommand{\topfigrule}{\vspace*{-1pt}%
\noindent{\color{cream}\rule[-\figrulesep]{\columnwidth}{1.5pt}} }

\newcommand{\botfigrule}{\vspace*{-2pt}%
\noindent{\color{cream}\rule[\figrulesep]{\columnwidth}{1.5pt}} }

\newcommand{\dblfigrule}{\vspace*{-1pt}%
\noindent{\color{cream}\rule[-\figrulesep]{\textwidth}{1.5pt}} }

\makeatother


\twocolumn[
  \begin{@twocolumnfalse}
\vspace{3cm}
\sffamily
\begin{tabular}{m{4.5cm} p{13.5cm} }

\includegraphics{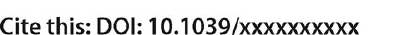} & \noindent\LARGE{\textbf{Anomalous behavior and structure of a liquid of particles interacting 
through the harmonic-repulsive pair potential near the crystallization transition}} \\
\vspace{0.3cm} & \vspace{0.3cm} \\

 & \noindent\large{Valentin Levashov,$^{\ast}$\textit{$^{ab}$} Roman Ryltsev,\textit{$^{cdb}$} Nikolay Chtchelkatchev\textit{$^{be}$}} \\

\includegraphics{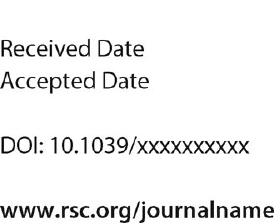} & \noindent\normalsize{A characteristic property of many soft matter systems is an ultrasoft effective interaction between their structural units. This softness often leads to complex behavior. In particular, ultrasoft systems under pressure demonstrate polymorphism of complex crystal and quasicrystal structures.  Therefore, it is of interest to investigate how different can be the structure of the fluid state in such systems at different pressures. Here we address this issue for the model liquid composed of particles interacting through the harmonic-repulsive pair potential. This system can form different crystal structures as the liquid is cooled. We find that, at certain pressures, the liquid exhibits unusual properties, such as the negative thermal expansion coefficient. Besides, the volume and the potential energy of the system can increase during crystallization. At certain pressures, the system demonstrates high stability against crystallization and it is hardly possible to crystallize it on the timescales of the simulations. 
To address the liquid's structure at high pressures, we consider the scaled pair distribution function (PDF) and the bond-orientational order (BOO) parameters. The marked change happening with the PDF, as pressure increases, is the splitting of the first peak which is caused by the appearance of non-negligible interaction with the second neighbors and the following rearrangement of the structure.
Our findings suggest that non-trivial effects, usually explained by different interactions at different spatial scales, can be observed also in one-component systems with simple one-length-scale ultrasoft repulsive interactions.} \\

\end{tabular}

 \end{@twocolumnfalse} \vspace{0.6cm}

  ]

\renewcommand*\rmdefault{bch}\normalfont\upshape
\rmfamily
\section*{}
\vspace{-1cm}



\footnotetext{\textit{$^{a}$~Technological Design Institute of Scientific Instrument Engineering, 630055, Novosibirsk, Russia. E-mail: valentin.a.levashov@gmail.com}}
\footnotetext{\textit{$^{b}$~Institute for High Pressure Physics, Russian Academy of Sciences,108840, Moscow (Troitsk), Russia}}
\footnotetext{\textit{$^{c}$~Institute of Metallurgy, UB RAS, 620016, 101 Amundsen str., Ekaterinburg, Russia}}
\footnotetext{\textit{$^{d}$~Ural Federal University, 620002, 19 Mira str,, Ekaterinburg, Russia.}}
\footnotetext{\textit{$^{e}$~Moscow Institute of Physics and Technology,141700, Institutskiy per.9, Dolgoprudny, Moscow Region, Russia}}


\section{Introduction}\label{s:intro}

The behavior of liquids is still poorly understood in comparison to the behavior of crystals.
This concerns, in particular, the structural and dynamical behavior of the liquids, 
as becomes especially evident in considerations of the behavior 
of supercooled liquids and the phenomenon of the glass transition.
One fundamental question is about the origin of the glass transition--is it 
driven by the structural changes or is it of purely dynamic
origin \cite{Tanaka2018,Royall2015,Chandler2010,Gokhale2016}?
Most related studies over the period of time larger than a century
were focused on atomic systems and model systems designed 
to model atomic systems \cite{Tanaka2018,Royall2015,Chandler2010}.
A characteristic feature of the atomic systems is a strong 
repulsion between the particles at small separation distances.

Developing abilities to design complex macromolecular and nanoparticle systems raised interest
in their simulations \cite{Gokhale2016,Kleman20031,Lang20001,Louis20001,Likos20011,Likos20012,Likos20021,Likos20061,Malescio20071,
Malescio20081,Frenkel20091,Prestipino20091,Saija20091,Malescio20111,Berthier20101,Zamponi2011,LuZY20111,Mohanty20141,Xu20141,Xu20151,Bagchi20181,Cipelletti2018}.
Model interactions in such systems, of course, can be quite different from typical interatomic interactions. 
One characteristic property of many soft matter systems is a finite repulsion even at vanishingly 
small separation distances between their structural units, i.e., 
many soft matter systems are ultrasoft. \cite{Lang20001,Louis20001,Likos20012,Likos20021}.

However, some properties of the systems composed of large molecules closely resemble certain phenomenons observed in the atomic systems.
This is related, in particular, to the phenomenons of the glass transition and jamming.
The similarities in the behaviors of these different systems open a possibility to test if 
the ideas developed for one type of systems are general enough to be valid for the systems 
of another type \cite{Gokhale2016,Cipelletti2018,Zamponi2011,Xu20141}.
One model system allowing to address these issues is the system of particles interacting 
through the harmonic-repulsive pair potential.
This and the other closely related potentials describe qualitatively ultrasoft effective repulsion between
globular micelles, microgels, starlike polymer solutions and other similar structural 
units of soft matter systems 
\cite{Berthier20101,Zamponi2011,LuZY20111,Xu20141,Xu20151,LuZY20111,Xu20181,
Lang20001,Likos20011,Frenkel20091,Prestipino20091,Saija20091,Zamponi2011,Denton20161}.

In our previous publication, we studied crystalline structures that form in the one-component 
system with the interaction between the particles described by the harmonic-repulsive
pair potential (HRPP) \cite{Levashov20161,Lokshin20181}.
It was observed that at different densities the system crystallizes into different crystal structures.
The formation of some of these structures, sometimes quite complex,
has not been anticipated previously \cite{Levashov20161,Lokshin20181,LuZY20111,Prestipino20091,Frenkel20091}.
Since in this system rather different crystalline structures can form, 
it is natural to expect that the structure of the liquid composed of
the harmonic-repulsive (HR) particles also can be quite different at different pressures.
In our view, it is of interest to investigate in a systematic way how
the structure of the HRPP liquid varies as pressure increases.
In our previous investigation, performed in
the NVT ensemble, we also observed that at some densities
the liquid exhibits remarkable stability against crystallization on cooling.
It is of interest to verify if the liquid exhibits
this behavior also in the NPT simulations.

In the present investigation, we consider a wider range of pressures as well 
as lower temperature ranges than in some of the previous investigations.
Besides, some of the methods that we apply were not used in the investigations 
of this system or in the ranges of parameters that we discuss here.

In this paper, we address how the structure of the liquid composed of particles interacting 
through the HRPP depends on pressure. In the investigations of the liquid's structure, 
we devote special attention to the pressures at which the liquid is stable against crystallization.

In our investigations of the liquid's structure, we are especially interested in
those structural changes which can not be described by simple rescaling of 
the interparticle distances.
To study this, we use two different methods to investigate the liquid's structure:
the scaled pair density function (PDF) and the bond-orientational order parameters.
The term ``scaled" here signifies that the PDF is properly scaled to address
the changes in the liquid structures which go beyond the simple rescaling of 
the interparticle distances.

We show that different non-trivial effects, such as the splitting of the first PDF peak,
high crystallization stability and water-like anomalies, which are usually observed
in systems with several different length scales in the interaction potentials, can also be observed
in simple single-component systems with one-scale ultrasoft repulsive interactions.

The paper is organized as follows. In section \ref{sec:methods}, the details of our computer simulations are described.
In section \ref{sec:general-prop}, the results on the general macroscopic properties of the system are presented.
Section \ref{sec:structure}, is devoted to the description of the structure of the system.
There, at first, we describe the structure of the liquid from the perspective of the scaled pair density function (SPDF).
After that, we address the structure with the bond-orientational order parameters (BOOP).
We conclude in section \ref{sec:conclusion}.

\section{Methods}\label{sec:methods}

The harmonic-repulsive pair potential used in our MD simulations
has the form:
\begin{equation}
  u(r)  =
  \begin{cases}
    \epsilon \left(1-\frac{r}{\sigma}\right)^2, & \text{if $r \leq \sigma$} \\
     0, & \text{if $r > \sigma $} \\
  \end{cases}\label{eq:hrp}
\end{equation}
In our simulations and in the description of our results further in the paper we measure energy in the units of $\epsilon$,
distance in the units of $\sigma$, and time in the units of $\tau = \left(m\sigma^2/\epsilon\right)^{1/2}$.

We used the LAMMPS molecular dynamics package to generate the liquids' structures at
different pressures and temperatures \cite{Plimpton1995,lammps}.
The Nose-Hoover non-Hamiltonian equations have been used to generate the coordinates
and velocities of particles (via the ``npt" and ``iso" commands within the LAMMPS).

Practically all results reported in this paper have been obtained on the system containing 8000 particles.
Some of the obtained results were compared with the results obtained on the system consisting of 65000 particles.
From these comparisons, which we do not discuss here, we concluded that there are essentially
no size effects in the results which we discuss in this paper.

The used value of time step at $T>0.010$ was $\delta t = 0.001\tau$,
while at $T<0.010$ the used value of time step was $\delta t = 0.010\tau$.
For $T<0.010$ the used value of the Nose-Hoover time-parameter used for the temperature
equilibration within the LAMMPS was $1\tau$, i.e., 100 time steps, while the used value of
the time-paramer for the pressure equilibration was $10\tau$,  i.e., 1000 time steps.
These are the recommended values for these parameters \cite{lammps}.

Initially, we generated the system as the FCC lattice at a very low density of  $\rho=0.04$.
Then the system was melted and equilibrated at $T=0.015$.
After the equilibration (which happens very fast at $T=0.015$),
the system has been cooled at $P=0.020$ down to $T=0.010$ which is still 
above any observable crystallization temperature for this system.
Then, at $T=0.010$, we increased the pressure from $P=0.020$ to $P=8.0$.
For $P<1.0$
the ``restart" files have been saved with the step in pressure $\Delta P = 0.05$, while for $P>1.0$
the ``restart" files have been saved with the step in pressure $\Delta P = 0.10$.
Starting from thus obtained restart files the systems have been equilibrated at all pressures at $T=0.010$.
At this high temperature the equilibration time at all pressures is smaller than $100 \tau$,
as can be judged from the dependence of potential energy on time.

Then we cooled the system(s) at different constant pressures.
The typical initial cooling rate used to observe crystallization was $10^6$ time steps per $\Delta T = 0.001$.
This cooling rate, for $P>1.0$, often was not sufficiently slow to observe crystallization.
Therefore, for $P>1.0$ we varied the cooling rate in the interval between $10^6$ and $10^7$ time steps per $\Delta T = 0.001$.
For most pressures these rates were sufficient to observe crystallization.
However, for certain pressures we did not observe crystallization even with the cooling rate $10^8$ time steps per $\Delta T = 0.001$.
In these cooling runs we, in particular, monitored how the mean square displacement of the particles depends on time and temperature.
From the dependence of the potential energy on temperature and time the value of the
glass transition temperature (where applicable) can be roughly evaluated.
These data, as expected, were in agreement with the results obtained from the monitoring of
the dependence of the mean square particle displacement on time and temperature.
In this paper we do not discuss the dynamic data.
Here, however, we eventually make references to the estimated glass transition temperatures.
These estimates come from the just discussed monitoring of the potential energy and the mean square displacement.

The liquid structures for the analysis have been collected after the equilibration at every considered value
of pressure and temperature.
The typical equilibration time at low temperatures of the liquids was $10^6$ time steps, i.e. $10^4 \tau$.
The structural configurations at low liquid temperatures, usually have been saved with
the time intervals $10^5$ or $2\cdot 10^5$ time steps.
During these intervals the mean square displacement usually increases by
more than $\sigma$ (the diameter of the particles).
Thus we can assume that these time intervals are sufficient to generate independent configurations.
The statistical averaging in most cases has been performed on 100 independent configurations.
The same structures have been used to produce various probability
distributions (PDs) presented further in the paper.

The analysis of the generated structures, in all discussed cases,
has been made with the self-made programs.
An additional analysis of the structures, with respect to
the bond-orientational order (BOO) parameters, has been performed
with the program ``Ovito" \cite{ovito1,ovito2}.
References to these results are made in a few places of the paper.

\section{Some general thermodynamic and structural properties of the liquid}\label{sec:general-prop}

In this section we describe some general macroscopic results obtained on the studied system(s).
These results are useful for the understanding of the data presented in further sections.
Due to a large amount of the accumulated data, we will show here, eventually, some illustrative
examples of the obtained results and then, in a different representation, the summary of all obtained data.

\begin{figure}
\begin{center}
\includegraphics[angle=0,width=3.5in]{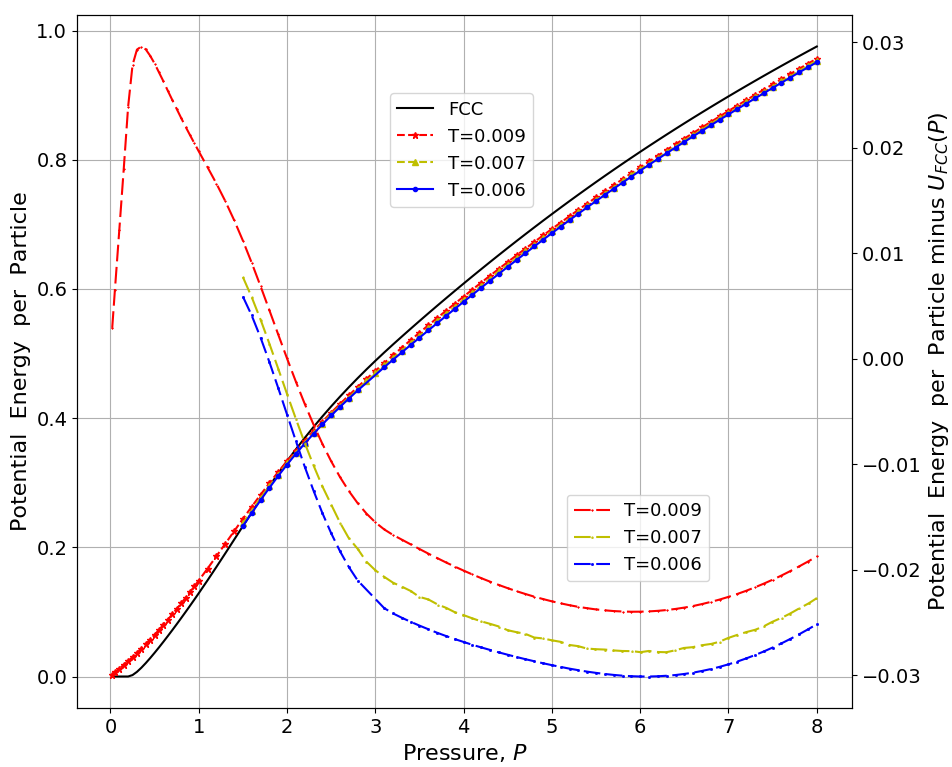}
\caption{The monotonically growing solid black curve shows the dependence
of the potential energy (PE) per particles, $U_{FCC}(P)$, of the FCC lattice on pressure at $T=0$.
The value of the pressure for the FCC lattice has been determined from the relation $P=-dU/dV$.
The other monotonically growing curves show the dependencies of the PEs of the liquid 
on pressure at selected temperatures.
The left $y$-axis describes the values of $U_{FCC}(P)$ and the PEs of the liquids.
The other three non-monotonic curves show the dependencies on the pressure of the differences 
between the PEs of the liquids and the PE of the reference FCC lattice at zero temperature.
The right $y$-axis corresponds to the values of these differences.
Note that at $P > 2.0$ the PEs of the liquids at non-zero temperatures
are smaller than the PE of the ideal FCC lattice at the same pressure.
}\label{fig:PE-minus-FCC}
\end{center}
\end{figure}

Figure \ref{fig:PE-minus-FCC} shows the dependence of the potential energy (PE) of
the liquid on pressure at selected temperatures.
These temperatures are not significantly above the (observable) melting temperatures.
In the following, as we make references to the melting temperatures,
we do not mean the true melting temperatures of the systems whose determinations
requires special efforts \cite{LuZY20111,Frenkel20091,Prestipino20091}.
In particular, in order to determine the true melting temperatures at different pressures,
it is necessary to know into which structures the liquid crystallizes at these pressures.
In our view, at present, there is no full clarity with respect
to this issue \cite{Levashov20161,LuZY20111,Frenkel20091,Prestipino20091}.
Thus, before addressing the true phase diagram of the studied system,
it is necessary to develop a better understanding of the structural properties of the system.
This is one of the purposes of this paper.
Thus, in our further discussion, as we use the term ``the melting temperature",
we mean only the melting temperature that we observed in our straightforward simulations.
In this context, we note that, from the thermodynamic perspective, 
we do not know if the liquids that we study are in equilibrium states or if they are in metastable supercooled states. 
This is also related to the regions of stability for the glass and crystalline states which we discuss in the
paper.

The choice of temperatures in Fig. \ref{fig:PE-minus-FCC} is dictated by the following observations.
At $T=0.008$ we already observed crystallization of the liquid at
pressures $P < 1.5$ into the HCP, FCC, and BCC lattices.
At $P \approx 2.5$ crystallization has been observed at $T \approx 0.006$.
At higher studies pressures, $2.5 < P < 8.0$, we observed crystallization in
the approximate interval of temperatures $0.004 < T < 0.006$.
See also the further discussion of Fig. \ref{fig:expansion-coeff-1}.

\begin{figure}
\begin{center}
\includegraphics[angle=0,width=3.5in]{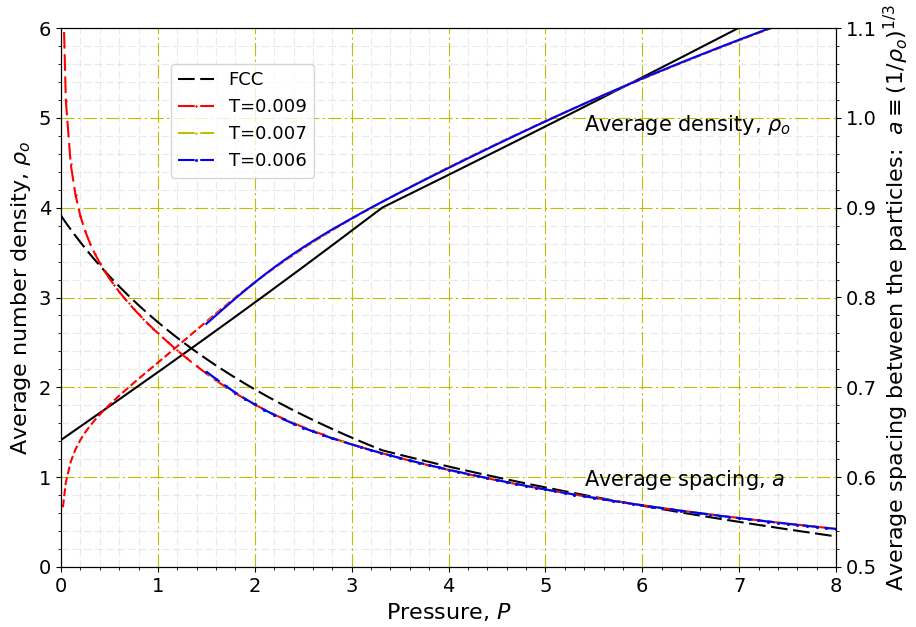}
\caption{The monotonically increasing curves describe the dependence of the average particles' 
number density, $\rho_o$, on pressure, $P$, for the ideal FCC lattice and the liquids at selected temperatures.
The results for the liquids at different temperatures essentially coincide.
The left $y$-axis shows the value of the density.
In drawing the parallel with the case of hard spheres,
it can be assumed that the packing fraction, $\phi$,
is related to the density via $\phi =(\pi \sigma^3/6)$.
If, as in our case, it is assumed that $\sigma=1$, then $\rho_o=1,\;2,\;3,\;4,\;5,\;6$
correspond to $\phi \approx 0.52,\;1.05,\;1.57,\;2.09,\;2.62,\;3.14$.
The monotonically decreasing curves show the dependencies on 
the pressure of the average spacing between the particles,
$a \equiv (1/\rho_o)^{1/3}$.
The value of the pressure for FCC lattice has been 
obtained from $PdV = -dU$.
}\label{fig:rho-vs-P}
\end{center}
\end{figure}

In Fig. \ref{fig:PE-minus-FCC} we also show with the solid black curve how the energy of
the ideal FCC lattice depends on pressure.
In order to find the relation between the density of the ideal FCC lattice and its pressure at zero temperature
we calculated the energy of the ideal FCC lattice as a function of density or the system's volume.
The zero-temperature pressure was found then from the relation $dP = -dU/dV=\rho^2(dU/d\rho)$.
Note in Fig. \ref{fig:PE-minus-FCC} that at $P < 2.0$ the energy of the ideal FCC lattice is
smaller than the energy of the considered liquids,
while for $P > 2.0$ the energy of the ideal FCC lattice is larger than the
energies of the considered liquids at non-zero temperatures.

In Fig. \ref{fig:PE-minus-FCC}, the changes in the PE of the liquid,
as its temperature is reduced at constant pressure,
are not noticeable on the presented scale of the left $y$-axis.
In order to show how the energy of the liquids depends on temperature,
we also present in Fig. \ref{fig:PE-minus-FCC} the differences between the PEs
of the liquids and the PE of the ideal FCC lattice at zero temperature and
at the same pressure as the pressure of the liquid.
The right $y$-axis addresses the values of these differences.

Figure \ref{fig:rho-vs-P} shows (the left $y$-axis) the dependence of the average particle
number density, $\rho_o$, on pressure at selected temperatures.
It also shows how the mean spacing between the particles, $a \equiv \rho_o^{-1/3}$,
depends on pressure (right $y$-axis).

Note from the figure that as the pressure increases from $P=1$ to $P=7$
the density changes from $\rho \approx 2.2$ to $\rho \approx 5.8$,
i.e., as pressure increases seven times the density changes $\approx 2.64$ times.
This behavior is noticeably different from what can be expected in the systems
with strong repulsion at small separations, i.e., in the liquid systems with strong repulsion
one would not expect such a strong dependence of the density on pressure as pressure increases less than 10 times.
Instead, in the liquid systems with strong repulsion, in order to change the average spacing
between the particles even by $2\%$ it might be necessary to increase pressure a 1000 times, as in the
case of water \cite{FineRA1973}.
As another comparative example, we can consider liquid iron in
the earth core conditions, i.e., at $T \approx 5000$ K.
In this situation, as pressure changes from $\approx 100$ GPa to $\approx 350$ GPa,
i.e., increases approximately 3 times
the volume of the systems changes by $\approx 10\%$ \cite{Ichikawa2014}.

In our NPT simulations we slowly cooled the liquid at different (constant) pressures.
For each pressure the volume of the system per particle and the potential energy of the system per particle
have been saved with the time step between $0.1\tau$ and $1.0\tau$.
The choice of the time interval for saving the data
depended on the cooling rate which was adopted to observe crystallization.
At some pressures, it was necessary to cool the liquid very slowly to observe crystallization.
For these slow cooling rates we were choosing larger time intervals for saving the data.

\begin{figure}
\begin{center}
\includegraphics[angle=0,width=3.3in]{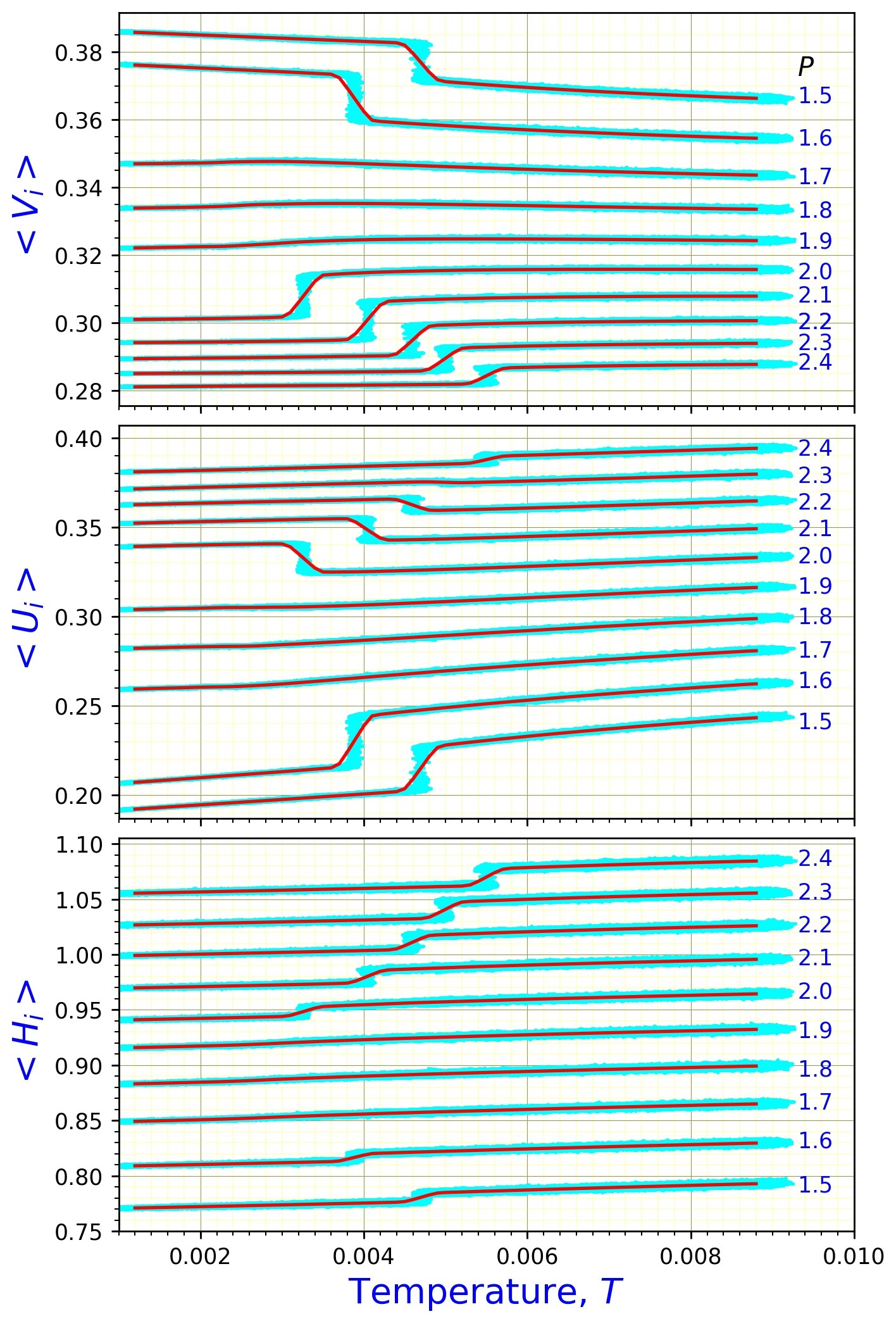}
\caption{
The top panel shows the dependencies of the volume per particle on temperature for the selected pressures.
The cyan curves in the figure show the data obtained directly from the simulations, 
while the red curves show the smoothened curves obtained through the least 
square fit procedure, as described in the text.
Note that for pressures $P=1.5$ and $P=1.6$ the volume per particle
increases as the liquid is cooled from $T = 0.9$ down
to the observed crystallization temperatures.
Thus, in this range of temperature and for the discussed pressures,
the constant pressure temperature expansion coefficient is negative for the liquid.
Also, note that for $P=1.5$ and $P=1.6$ the volume of the system increases as crystallization happens.
Finally, note that for pressures $P=1.7$, $P=1.8$, and $P=1.9$ we do not observe crystallization in the results of the simulations.
The middle panel shows the dependencies of the potential energy per particle on temperature for the selected pressures.
Note that for pressures $P=2.0$, $P=2.1$, and $P=2.2$ the potential energy increases in the process of crystallization.
The bottom panel shows the dependencies of the enthalpy per particle, $<H_i>\equiv <U_i> + P<V_i>$, on temperature for selected pressures.
Note that the enthalpy always decreases in the process of cooling and also always decreases in the process of crystallization.
}\label{fig:expansion-1}
\end{center}
\end{figure}

The examples of the raw data from the simulations are shown as the cyan curves in
Fig.\,\ref{fig:expansion-1}.
The chosen pressure interval, $1.5 < P < 2.4$, from which the data are shown, is actually of significant interest.
As follows from our previous work, 
at $P=1.5$ the low-temperature crystal structure of the system is the BCC lattice, 
while 
at $P=2.4$ the low-temperature structure is the $Ia\bar{3}d$ crystal \cite{Levashov20161}.
Thus, in the pressure interval $1.5 < P < 2.4$ a solid-solid transition is expected.

Furthermore, in the discussed pressure interval we also observed the formation of a rather complex
structure which we initially assumed to be a quasicrystal or a quasicrystalline approximant.
However, later we interpreted this structure as a rather complex crystal \cite{Levashov20161}.
In our previous investigations, we studied the system in the NVT ensemble and we studied the systems
containing $13500$ and $32000$ particles. 
In those systems, we observed the formation of the discussed complex structure. 
Recently, we repeated these simulations and in several different long runs and we again observed the formation of this structure.
It is of interest that the discussed structure forms much more readily in the smaller system of
$13500$ particles than in the larger system of $32000$ particles. 
This might be related to some commensuration effects.
In our current study, as we described in section \ref{sec:methods}, 
we mostly studied the behavior of the system containing $8000$ particles.
We also studied the systems in the NPT ensemble. 
In these NPT simulations, despite rather long cooling runs, 
we did not observe the process of crystallization at $P=1.80$ which
closely corresponds to the density $\rho_o = 2.904$, i.e., the density at which we observed the formation
of the complex structure in our previous work (see Fig. \ref{fig:rho-vs-P}) . 
Investigations of the formation of this complex structure is not the topic of this paper.
We note, however, that in Ref. \cite{NingXu2017} observations of the quasicrystalline structures in $2D$ 
in the system interacting through the HR potential have been reported. 
Moreover, it has been stated there in the supplementary materials that there, possibly, also
form quasicrystalline structure in $3D$.

Note in the top panel of Fig. \ref{fig:expansion-1} that 
for $P=2.0$, $P=2.1$, and $P=2.2$ 
the volume of the system increases as temperature decreases.
Then note also in the middle panel 
of Fig.\ref{fig:expansion-1} that for pressure from $P=1.6$ to $P=1.9$
the potential energy of the system increases
in the process of crystallization.
Finally, in the bottom panel of Fig.\ref{fig:expansion-1} we 
show the dependence of the enthalpy of the system on temperature.
It follows from the bottom panel that the enthalpy of the system 
always decreases in the process of crystallization.
Therefore, we can not conclude from the presented data if the crystallization 
process is driven by the enthalpy or by the entropy.

\begin{figure}
\begin{center}
\includegraphics[angle=0,width=3.5in]{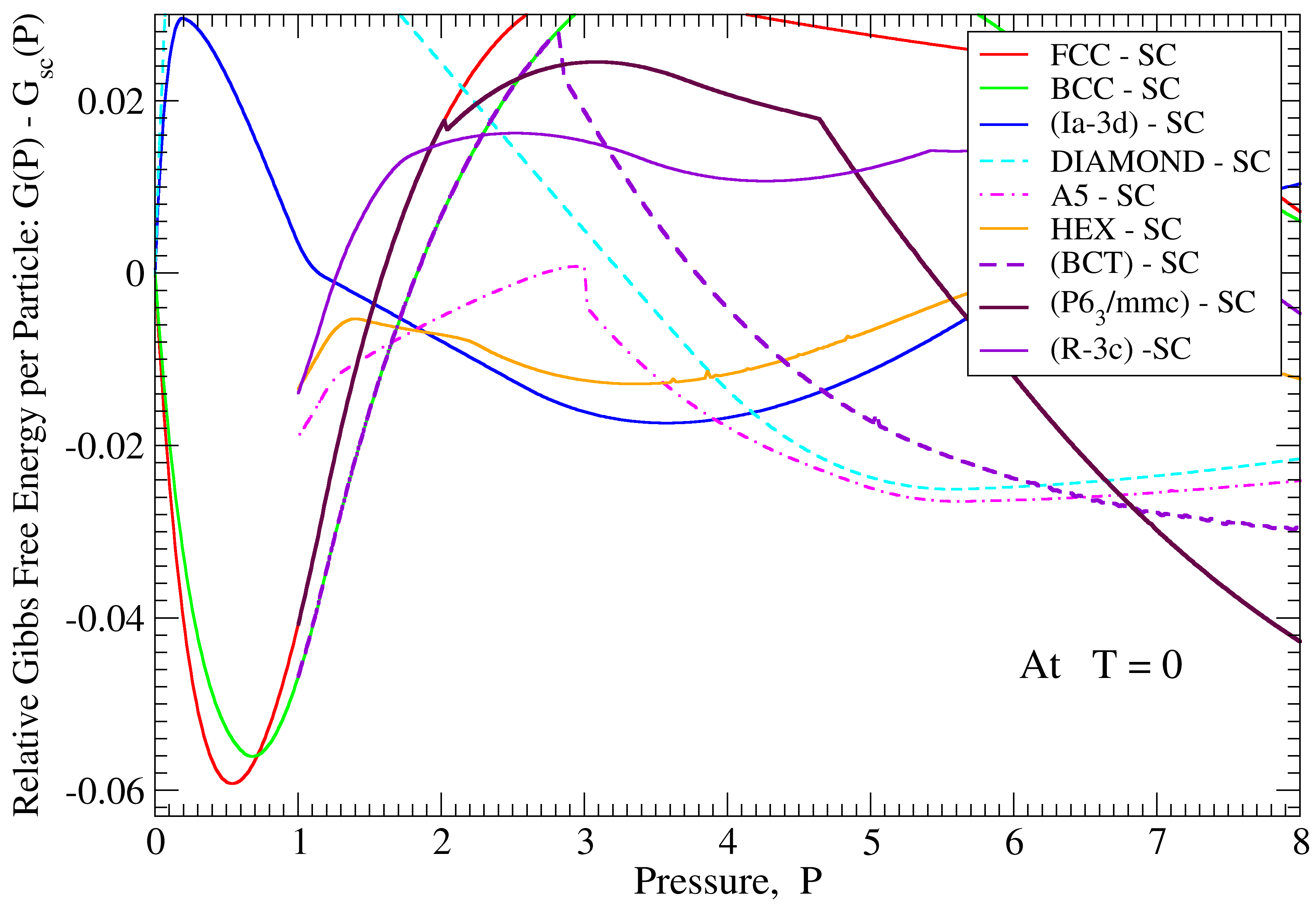}
\caption{
The dependencies on pressure at $T=0$ of the differences between the Gibbs free
energies (chemical potentials) for the selected lattices and the Gibbs free energy for the simple cubic lattice.
Note that at very low pressure the FCC lattice has the lowest value of the Gibbs free energy.
As pressure increases the BCC lattice becomes more stable than the FCC lattice.
As pressure increases further the $Ia\bar{3}d$ becomes the most stable
between the considered lattices.
Note, however, that we did not calculate the Gibbs free energy for the $C2/c$ structure whose
region of stability can be expected to occur between the regions of stability for the BCC and
the $Ia\bar{3}d$ lattices. On further increase of pressure the $A5$ (i.e., the distorted diamond)
structure becomes more stable than the $Ia\bar{3}d$ crystal structure.
Then the $P6_3/mmc$ lattice becomes the most stable. These results are presented in a concise form also in Table \ref{table:phase-regions}.
}\label{fig:gibbs0}
\end{center}
\end{figure}

\begin{figure}
\begin{center}
\includegraphics[angle=0,width=3.4in]{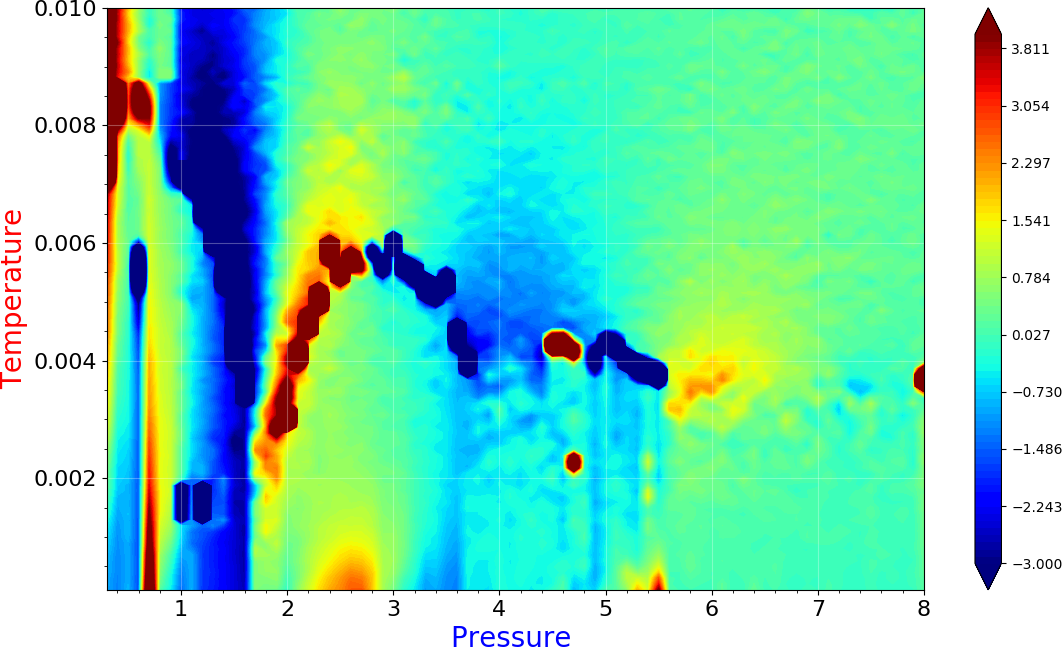}
\caption{
The dependence of the constant pressure temperature expansion coefficient on pressure and temperature.
The data have been accumulated in the constant pressure simulations as
the temperature of the liquid has been reduced.
The regions of high red intensity correspond to crystallization with an abrupt decrease in the systems' volume.
The regions of high blue intensity correspond to crystallization
with an abrupt increase in the systems' volume.
In the blue regions the temperature expansion coefficient is negative.
The figure could be considered as a crystallization ``phase diagram" obtained
from the direct cooling process and thus it should be compared with Fig.\ref{fig:gibbs0}.
}\label{fig:expansion-coeff-1}
\end{center}
\end{figure}

In order to gain further insight into the origin of such a complex behavior,
it is useful to consider the zero-temperature phase diagram of the system which
is based on our previous investigation of the crystalline
structures which form in the harmonic-repulsive system.
For this we in Ref. \cite{Levashov20161} calculated the Gibbs free energies of
the structures observed in our simulations at zero temperature.
These results (in a slightly modified form) are shown in Fig.\ref{fig:gibbs0}
where, for the clarity of presentation, we plot the Gibbs free energies
of the considered lattices with respect to the free energy of the simple cubic lattice.

As follows from Fig.~\ref{fig:gibbs0}, at $P \sim 1.80$, there are several lattices which
have almost identical values of the Gibbs free energy at zero temperature.
Thus, at non-zero temperatures, the system might be ``frustrated" with
``choosing" the lattice into which it should crystallize.
This can cause the stability of the liquid with respect to crystallization,
which we observed
in Fig. \ref{fig:expansion-1}
in the curves corresponding to $P=1.7$, $P=1.8$, and $P=1.9$. 
A similar mechanism for the stability against crystallization was recently proposed 
in systems with core-softened pair potentials \cite{Ryltsev2013}. 
In these systems, a strong competition between interparticle 
scales was generated by two-scale nature of the potential.
In our case, the competition is, probably, 
caused by ultrasoftness of the interatomic potential.

In table \ref{table:phase-regions} we summarize the results on the regions of stability of the lattices that we observed in
Ref.\cite{Levashov20161}.
\begin{center}
\begin{table*}
\begin{tabular}{| c | c | c | c | c | c |} \hline
$0.3 < P < 0.7$  & $0.7 < P < 1.7$  & $1.7 < P < 3.9 $  &   $3.9 < P < 6.6$  & $6.6 < P < 6.9$    & $6.9 < P < 8.0$   \\ \hline
$FCC$            & $BCC$            & $Ia\bar{3}d$      & $A5$        &$BCT$            & $P6_{3}/mmc$          \\ \hline
\end{tabular}
\caption{The calculated regions of stability at zero temperature for the lattices observed in MD simulations. The results
presented in the table follow from Fig.\ref{fig:gibbs0}.
}
\label{table:phase-regions}
\end{table*}
\end{center}

In our further processing of the data, similar to those shown
in Fig. \ref{fig:expansion-1}, we were choosing
the temperatures with the step in temperature $\Delta T = 0.0001$.
Then for temperature intervals $(T-0.0002 : T+0.0002)$ we used the linear least
square fitting procedure to determine the slopes of the curves that describe 
the dependence of the volume per particle on temperature in the top panel of Fig. \ref{fig:expansion-1},
the dependence of the potential energy per particle on temperature in the middle panel of Fig. \ref{fig:expansion-1},
and the dependence of the enthalpy on temperature in the bottom panel of Fig. \ref{fig:expansion-1}.
The corresponding average values are shown as red curves in Fig. \ref{fig:expansion-1}.
The slopes of the red curves in the top panel of Fig. \ref{fig:expansion-1} (divided by the volume per particle)
provide the values of the constant pressure expansion coefficient, $\beta_p \equiv (1/V)(dV/dT)_p$.

The dependence of the constant pressure expansion
coefficient on the pressure and temperature of the system
is summarized in Fig. \ref{fig:expansion-coeff-1}.
In the yellow and reddish areas of the figure the constant pressure expansion coefficient is positive,
while in the blue areas the constant pressure expansion coefficient is negative.
The crystallization in the simulations has been observed at the values of
the pressure and temperature at which we see in the figure
very intense red or blue colors.
The bright red lines-regions correspond to the values of $P$ and $T$ at which crystallization happens with
a decrease of the systems' volume, while intense blue regions correspond to the crystallization with an increase of the system's volume.
Thus, effectively, Fig. \ref{fig:expansion-coeff-1} represents a crystallization diagram of the system
which can be compared with Fig. \ref{fig:gibbs0}.

Note that it follows from Fig. \ref{fig:expansion-coeff-1} that there also might 
happen solid-solid transitions on cooling at constant pressures. 
The signatures of these possible solid-solid transitions are the bright intensity spots at 
$[P \approx 0.6,\;\;T\approx 0.0055]$, 
$[P \approx 0.6,\;\;T\approx 0.0015]$, and 
$[P \approx 4.7,\;\;T\approx 0.0022]$.
Investigations of the possible solid-solid transitions are beyond 
the scope of this paper and we do not discuss this issue further.

It is important to note that both negative thermal expansion coefficient and the increase in
specific volume upon crystallization are the so called water-like anomalies,
which are observed in water as well as in different model systems mimicking its behaviour.
In particular, such effects are observed in systems with core-softened potentials where
competition between different spatial scales takes place \cite{Scala2001,Debenedetti1991}.
Here we observe the same effects in the system with one-scale potential
where the same competition between spatial scales is caused by softness of the potential.

It also follows from Fig. \ref{fig:expansion-coeff-1} that, at $P\approx 4.0$ and
$P > 6.0$, we did not observe crystallization on simulation timescales.

It is interesting, though rather predictable,
that the liquid exhibits stability against crystallization at pressures
which are close to the boundary values for the different crystal lattices.
It also should be noted that the formation of the quasicrystal-like structure,
which we reported in Ref. \cite{Levashov20161},
appear to happen exactly at the value of
the density which may not be allowed in the temperature-density phase diagram, i.e.,
in the phase-separation region between the two crystal structures.

\section{On the Structure of the Harmonic-Repulsive Liquid}\label{sec:structure}

Structural properties of the HRPP liquid at high temperatures and pressures
have been considered briefly through the pair density function in Ref. \cite{Xu20151}.
A discussion of the behavior of the first peak of the PDF at $P < 1.0$ has been presented in
Ref. \cite{Berthier20101}.
The other (already mentioned) references related to the structure of
the harmonic-repulsive system are Ref. \cite{Xu20141,Xu20151,LuZY20111,Berthier20101,Levashov20161}.

Our study is different from the previous investigations in that we consider the behavior of
the system in the broader range of pressure in comparison, for example, with Ref. \cite{Xu20141,Berthier20101}.
We also consider the behavior of the system at lower temperatures than in Ref. \cite{Xu20151}.
Besides, we utilize a rather straightforward approach in a hope to observe the behaviors 
which cannot be anticipated in advance,
such as negative temperature expansion coefficient already discusses in the previous section.
Here we are interested in the structural properties of the harmonic repulsive liquid at low temperatures,
i.e., at temperatures not significantly above the crystallization curves shown in Fig. \ref{fig:expansion-coeff-1}.
In particular, we are interested in the evolution of the structural changes which can
not be reduced to the simple rescaling on the inter-particle separations.

In our studies, we used two approaches.
In the first approach, we consider the scaled pair density function,
while in another approach we consider certain bond-orientational
order parameters at the selected pressures and temperatures.

\subsection{Pair Distribution Functions (PDFs)}\label{sec:pair-distr}

In order to address the structural changes of the liquid beyond the simple rescaling of the inter-particle distances,
we, besides considering the standard pair density function, $g(r)\equiv \rho(r)/\rho_o$ ,and the standard
pair distribution function, $G(r)\equiv 4\pi r \rho_o \left[g(r)-1\right]$, 
also considere their scaled analogues in which the distance is measured in terms of the average interparticle distance, 
$a$, defined through the average density $\rho_o = 1/a^3$:
\begin{eqnarray}
g_{a}(r/a) \equiv \frac{\rho(r)}{\rho_o},\;\;\;
G_{a}(r/a) \equiv 4\pi \left(\frac{r}{a}\right) \left[g_a(r/a) - 1\right].\;\;\;
\label{eq:pairdensityfunction}
\end{eqnarray}

Before discussing the structure of the HRPP liquid at relatively low temperatures and relatively 
high pressures, it is useful to present certain results for this system which allow putting our further results
in a broader context. 
As we already mentioned above, the major result concerning the behavior of systems 
with ultrasoft pairwise interactions is related to the
``clustering" vs. ``reentrant melting" behavior \cite{Likos20011}. 
In the same reference \cite{Likos20011}, there were made certain predictions concerning 
the structural behavior of such systems in the limit of high pressures and temperatures, i.e., in the
limit when the mean-field approximation (MFA) for the excess free energy can be adopted.
In particular, it was shown that, in the MFA limit, 
the structure factor $S(Q)$ for the system is determined 
by a simple combination of the density, temperature, and the
3-dimensional Fourier transform of the pair interaction potential, $\hat{\phi}(Q)$, 
i.e., $S(Q)$ is determined by $\left(\rho/T\right)\hat{\phi}(Q)$.
Furthermore, the expression for $S(Q)$ is such that in the limit of very large $(\rho_o/T)$ 
we have $S(Q) \sim \left[(\rho_o/T)\hat{\phi}(Q)\right]^{-1}$.
It was also demonstrated for several systems that the mean-field approximation becomes 
quite valid for the pressures and temperature which are not extremely high \cite{Likos20011,Likos20021}.
Since there is a direct correspondence between $S(Q)$ and $g(r) \equiv \rho(r)/\rho_o$, 
we should, of course, expect that $g(r)$ also depends only on the already mentioned combination $\left(\rho/T\right)\hat{\phi}(Q)$. 
Furthermore, it follows from the limiting expression for $S(Q)$ 
that in the limit of very large $(\rho_o/T)$ the periodicity of
$g(r)$ should be completely determined by the pair potential, i.e., 
it should not depend on the density.
In this respect, the situation for the HRPP (and the other $Q^{+}$ systems) 
is somewhat similar to the situation with the systems in which the clustering occurs--there
the distance between the clusters does not depend on 
the total number of particles in the limit of very large densities. 
The reason for the similar behaviors of both types of systems is the same.

\begin{figure}
\begin{center}
\includegraphics[angle=0,width=3.2in]{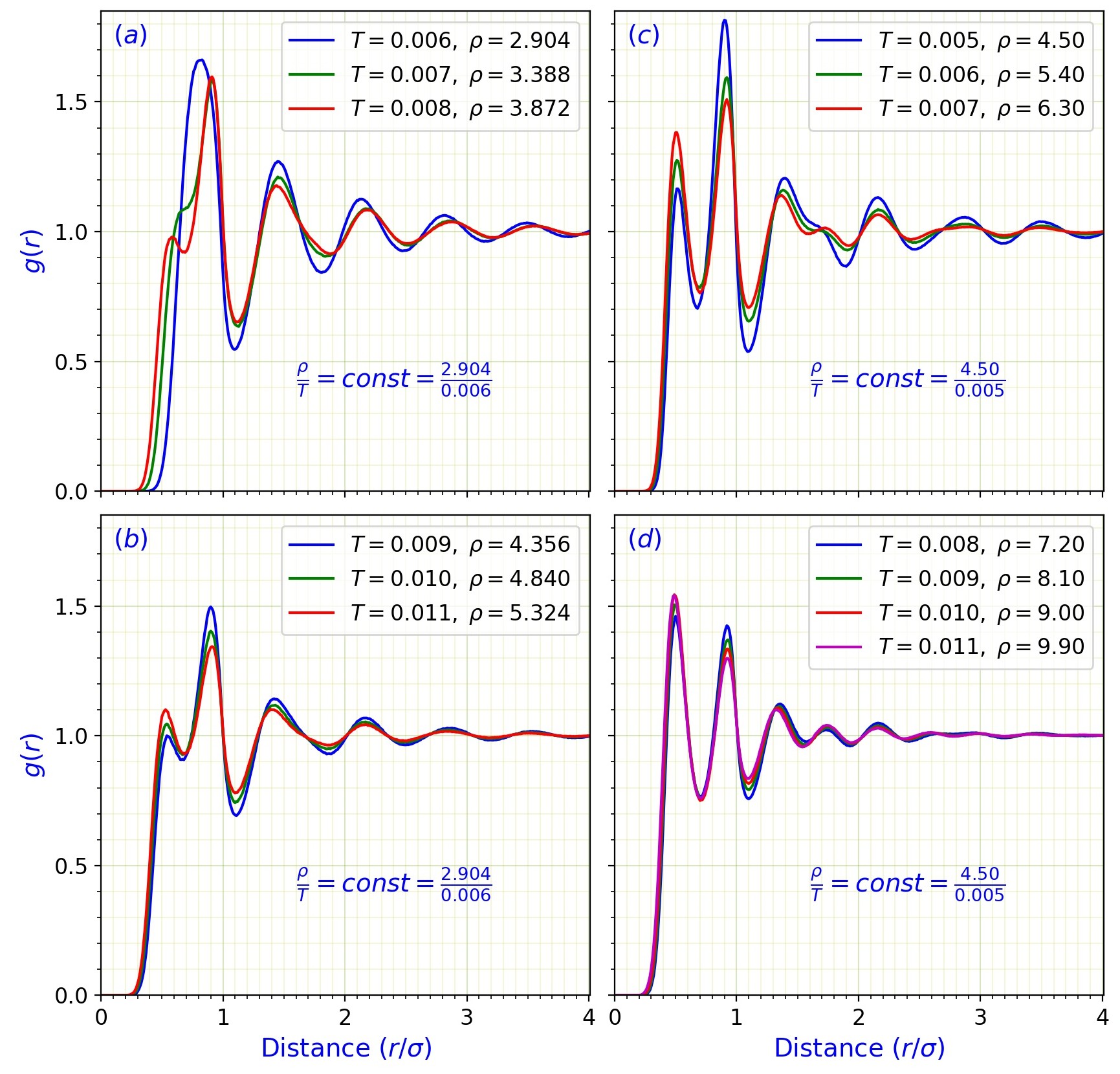}
\caption{
The pair density functions $g(r)\equiv \rho(r)/\rho_o$ for the liquid states for the selected densities
and temperatures. The curves in panels (a,b) correspond to the fixed ratio of $\rho/T = 2.904/0.006$, 
while the curves in panels (c,d) correspond to the fixed ratio of $\rho/T = 4.50/0.005$.
Note that the high pressure and high temperature curves in panels (b,d) appear to converge to some
limiting curves. On the other hand, the curves in panels (a,c) appear to be quite different. 
}\label{fig:rho-t-phi}
\end{center}
\end{figure}

In Fig. \ref{fig:rho-t-phi} we show the pair density functions, $g(r)$,
for the liquid states corresponding to the two different values of $(\rho/T)$. 
Thus, in panels (a,b) we show the results for the selected values of $\rho$ and $T$ such that $\rho/T = const=(2.904/0.006)$, 
while in panels (c,d) we show the results for the different values of $\rho$ and $T$ such that $\rho/T = const=(4.500/0.005)$. 
These ratios are of particular interests to us because at densities $\rho = 2.904$ and $\rho = 4.50$ the HRPP system exhibits remarkable stability against crystallization.
It follows from Fig. \ref{fig:rho-t-phi} that, as 
the density and temperature of the liquid increase in a way that their ratio remains constant,
the behavior of $g(r)$ converges to certain limiting curves, 
as expected from the MFA and Ref. \cite{Likos20011}.
It also follows from the curves presented in panels (a,c), 
that for the temperatures and densities addressed in (a,c) the MFA approximation is not appropriate.

We now turn our attention to the results obtained in the constant pressure simulations.

\begin{figure}
\begin{center}
\includegraphics[angle=0,width=3.1in]{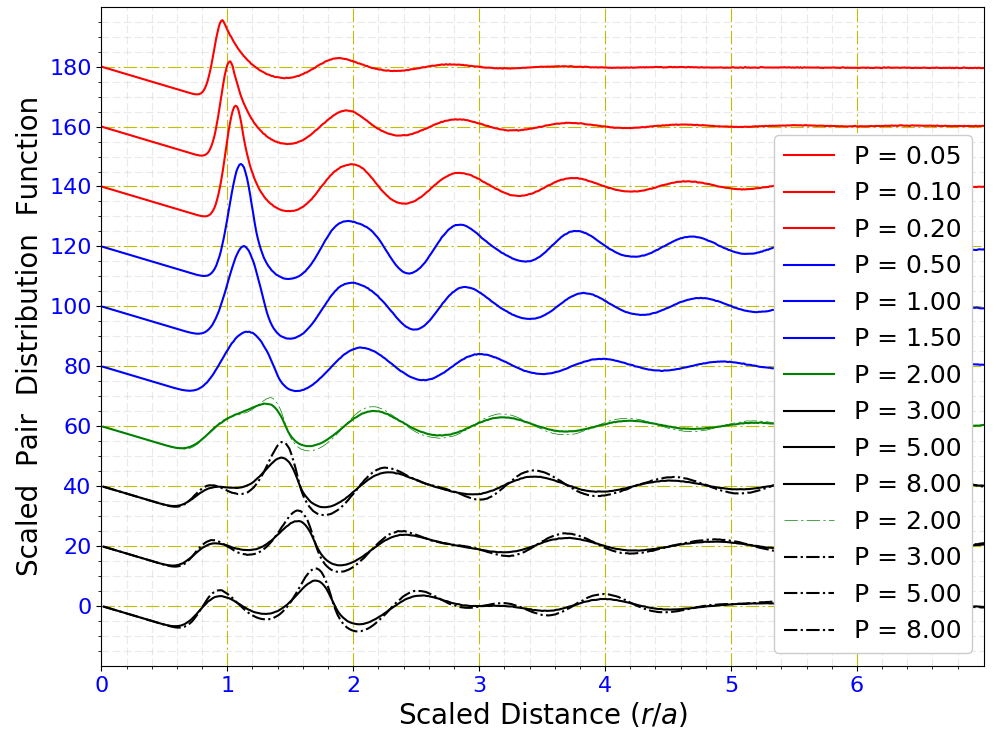}
\caption{The scaled pair distribution functions (SPDFs), defined in (\ref{eq:pairdensityfunction}),
of the liquid states at selected pressures at $T=0.009$ (solid curves) and at $T=0.006$ (dashed curves).
The curves corresponding to the lower pressures were offset vertically for the clarity of the presentation. 
Thus, each curve for the next higher pressure is shifted downward (by 20) with 
respect to the curve at the previous lower pressure.
}\label{fig:pdf-select-01}
\end{center}
\end{figure}

The scaled pair distribution functions (PDFs) calculated 
for $T=0.009$ and $T=0.006$, for the selected pressures, 
are shown in Fig. \ref{fig:pdf-select-01}.
The most marked change observed in the scaled PDF, 
as pressure increases, is the splitting of its first PDF peak 
into two sub-peaks (this behavior, of course, is observed in the unscaled PDF also).
Similar splitting is often observed in two-length-scale systems
where different bond lengths are generated by the special form of
the interparticle potentials, like the two attractive wells or
the repulsive shoulder, and so on \cite{Fomin2008,Ryltsev2013,Ryltsev2015,Ryltsev2017}.
Note, that the same effect was observed in star polymers solutions modeled by the ultrasoft 
potential consisting of two parts:
the logarithmic repulsion at $r<\sigma$ and the exponential Yukawa-like tail at $r>\sigma$, 
where $\sigma$ is the characteristic corona diameter \cite{Watzlawek1998,Likos20021,Likos20012}.
It was assumed previously that the existence of the crossover between the two parts 
of the interaction potential is mainly responsible for the PDF splitting.
Our results suggest that, in the case of the HRPP and, obviously, other similar potentials,
this effect can be caused simply by the shape of the potential,
which allows for the non-negligible interaction with 
the second neighbors (and may be even with the further neighbors).

Note that the positions of the peaks in the scaled PDF at larger distances shift to 
even larger distances as the pressure increases. Besides, at the largest pressure $P=8.0$,
we see the splitting of the second peak in the interval of distances
$2.1 < (r/a)< 3.5$.

\begin{figure}
\begin{center}
\includegraphics[angle=0,width=3.1in]{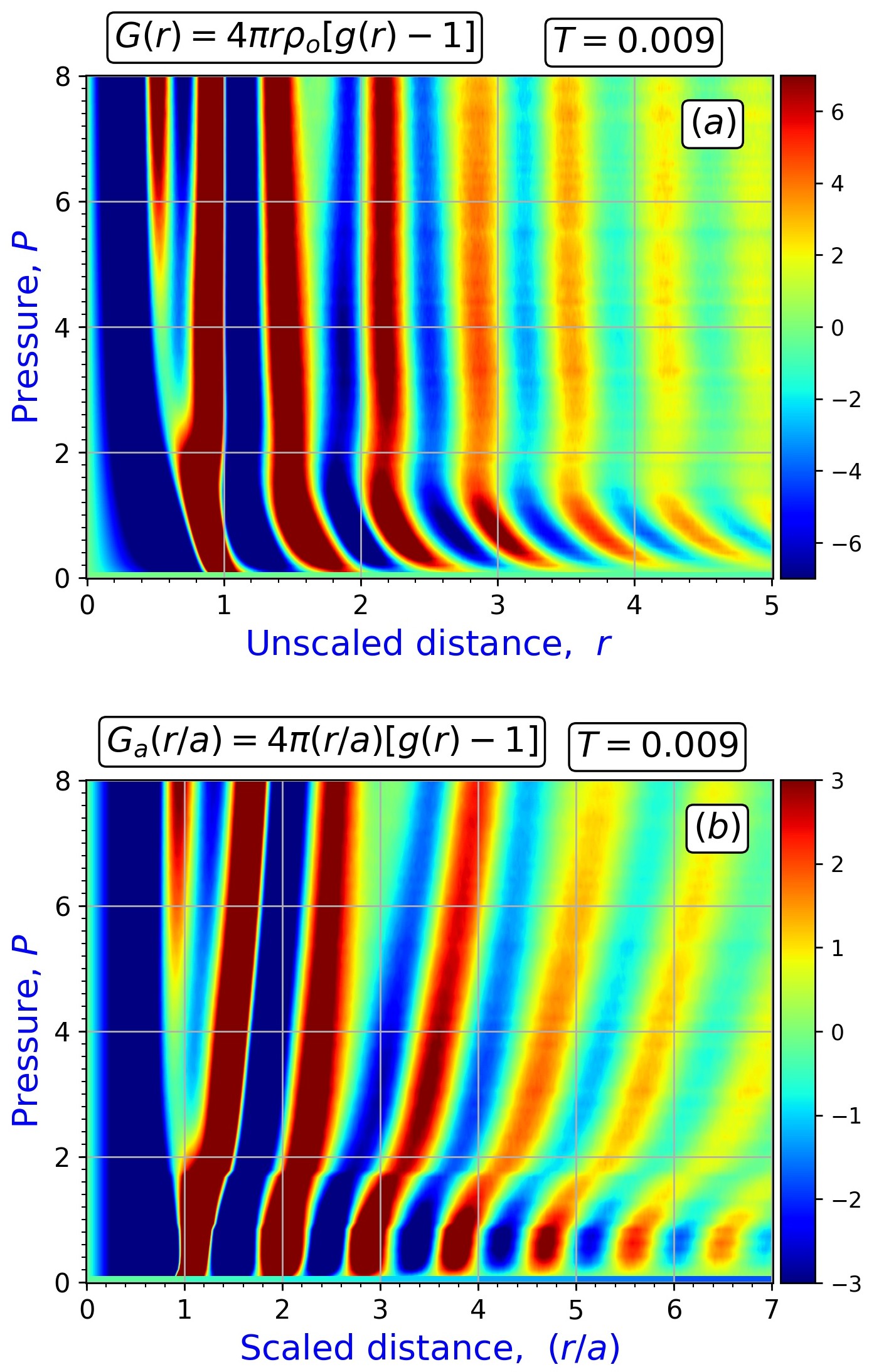}
\caption{Panel (a) shows the 2D contour plot of the unscaled PDF,
$G(r)=4\pi r \rho_o\left[g(r)-1\right]$ as a function of 
the unscaled distance, $(r/\sigma)$, at different pressures.
Note that for $P>2.5$ the positions of the peaks in PDF at distances beyond the second neighbors
do not exhibit pressure dependence.
Panel (b) shows the behavior of the scaled PDF, i.e., the behavior
of $G_a(r/a) \equiv 4\pi (r/a)\left[g(r)-1\right]$ 
as a function of the scaled distance $(r/a)$, 
where $a$ is the average separation between the particles,
$\rho_o = 1/a^3$. 
The difference between the top and bottom panels highlights the increase in the system density,
i.e., the decrease in $a$, as the system's pressure increases.
}\label{fig:spdf2d0x009-1}
\end{center}
\end{figure}

To grasp the changes which happen with the PDF, as the pressure increases, it is reasonable to consider also the 2D contour plots of the unscaled and scaled PDFs which are shown in Fig. \ref{fig:spdf2d0x009-1}.
As in Fig. \ref{fig:pdf-select-01}, we see in Fig. \ref{fig:spdf2d0x009-1}
that, as the pressure increases, the first peak splits into two.
In panel (a), corresponding to the unscaled PDF, 
the position of the outer shell, originating from the splitting of 
the first peak, remains in place at high pressures. 
The position of the internal shell, after the splitting at the intermediate 
pressures $2.0<P<5.0$  at high pressures also remains in place at $P>5.0$ 
The positions of the peaks in the PDF at large distances, $r>1.4$, also do not change, 
as the pressure increases beyond $P\approx 2.5$.
This independence of the peaks' positions on the pressure, at high pressures, 
follows from the fact that the dependence of $S(Q)$ on $Q$ 
is completely analogous to the dependence on $Q$ of $\hat{\phi}(Q)$,
i.e., it does not depend on the density.

The behavior of the scaled PDF is shown in panel (b) of Fig. \ref{fig:spdf2d0x009-1} . 
We see that the positions of the peaks in the scaled PDF exhibit a very clear pressure 
dependence that demonstrates that the density of the system markedly increases 
as the pressure increases and, correspondingly, the average distance between the particles, $a=\rho_o^{-1/3}$, decreases. 
Note that the behavior which we observe for the ultrasoft HRPP system is opposite 
to what is routinely observed for the systems with strong diverging repulsions at short distances.
Thus, in the systems with the strong repulsions, the positions of the peaks 
in the scaled PDF essentially should not exhibit any dependence on the pressure, 
while the positions of the peaks in the unscaled PDF should only weekly depend on the pressure.
 
Both panels of Fig. \ref{fig:spdf2d0x009-1} show that a pronounced change in 
the structure at all distances happens in the range of pressures $1.0<P<2.0$.
This change is more pronounced at larger distances than at smaller distances.
It is also possible to say that the structural changes propagate 
from the larger distances to the smaller distances as the pressure increases from $P \approx 1$ to $P \approx 2$.
Finally, around $P=2.0$, the structural changes result in the splitting of the first peak.
This change should be associated with a transition in which the interaction 
with the second neighbors becomes important.
Note that the inner shell, resulting from the splitting, has a smaller number 
of neighbors in it, in comparison, to the outer shell that resulted from the splitting.
Thus the observed splitting is different from the ``splitting" 
of the first shell observed in the BCC lattice.
Also, note that for $P > 8.0$ there might also happen the splitting 
of the second peak.


While we do not present here the 2D contour plot of the PDF at lower temperature $T=0.006$,
the obtained data show that the splitting of the first peak, as expected, becomes more pronounced at this lower temperature.
The data from $T=0.006$ also show that there indeed happens the splitting 
of the second peak at $P>7.5$ into the main part located at shorter distances 
and a smaller part located at larger distances.

\begin{figure}
\begin{center}
\includegraphics[angle=0,width=3.2in]{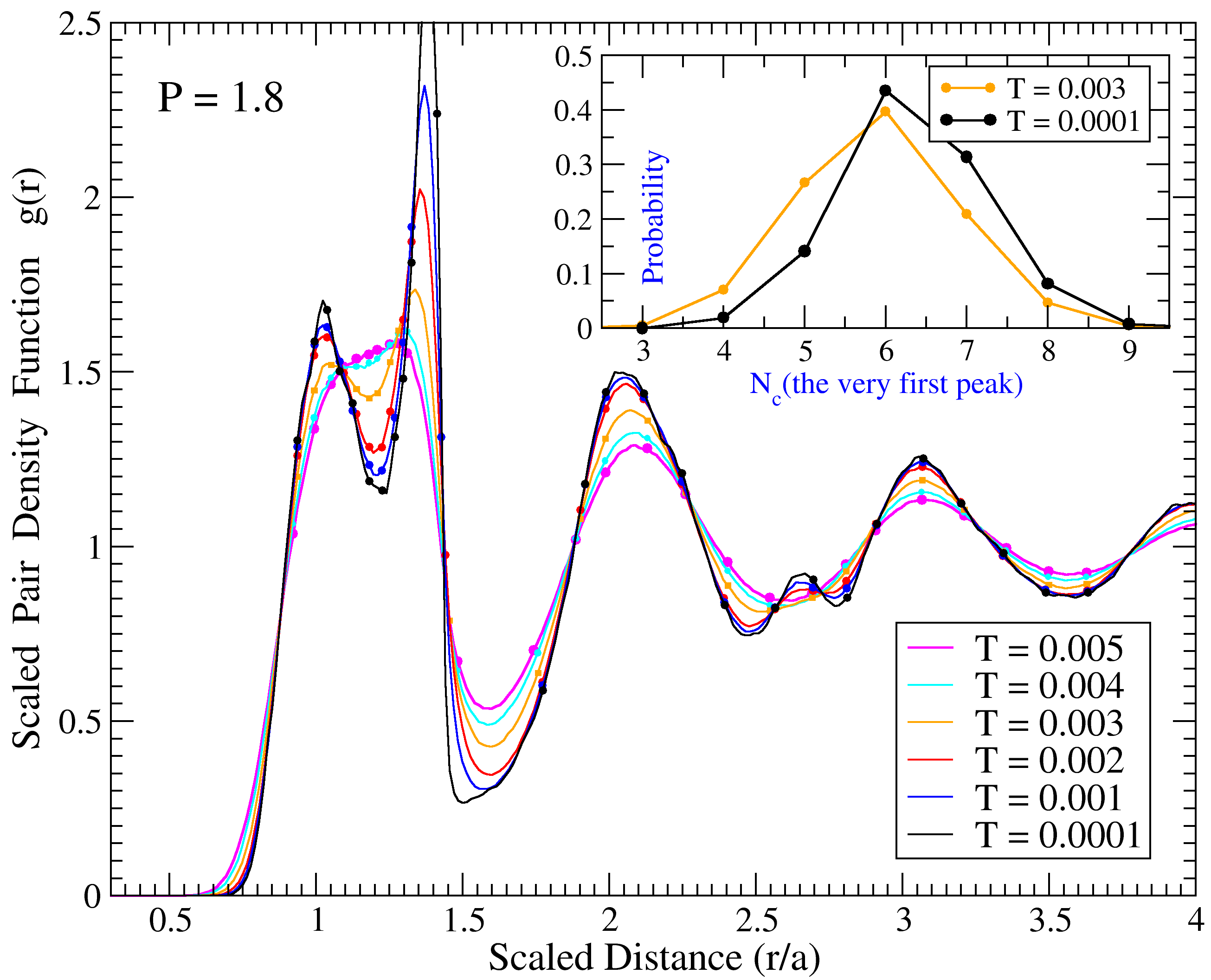}
\caption{The scaled pair density functions (PDFs), $[\rho(r/a)/\rho_o]$ vs. $(r/a)$, of
the liquid and glass states at $P=1.80$.
The states at lower temperatures were obtained by cooling the higher temperature
states at constant pressure.
The points on the curves correspond to the scaled distances at which the proper
integrals of the radial density lead to the coordination numbers 1, 2, 3, 4, 5, 6, 7, 8, 9, 10,
15, 20, 25, 30, 40, 50, 60, 70, 80, 90, 100, 120, 140, 160, 180, 200.
Note that in the lower temperature glass states, i.e., at $T \leq 0.002$ there
develops an additional small peak at $(r/a) \approx 2.67$.
The inset shows the distributions of the coordination numbers for the particles.
Here the coordination number is defined through the position of the minimum after the pre-peak.
Thus, most of the particles have the coordination numbers 5, 6, or 7.
}\label{fig:rhor-points-P-1x8}
\end{center}
\end{figure}
\begin{figure}[t]
\begin{center}
\includegraphics[angle=0,width=3.2in]{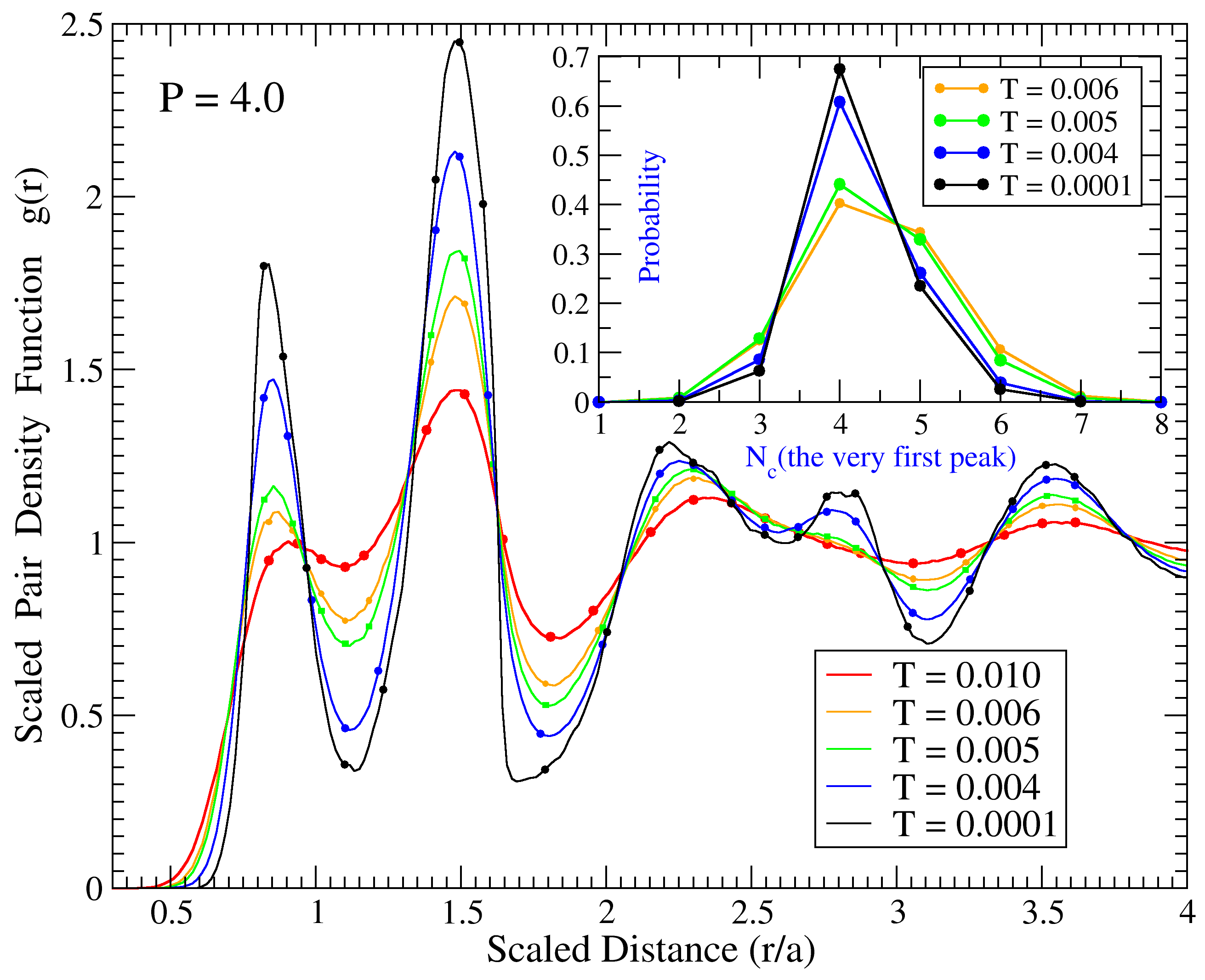}
\caption{
The scaled pair density functions (PDFs) of the liquid and glass states at $P=4.0$.
The states at lower temperatures were obtained by cooling
the higher temperature states at constant pressure.
The points on the curves correspond to the distances at which the proper
integrals of the PDFs lead to the coordination numbers 1, 2, 3, 4, 5, 10,
15, 20, 25, 30, 40, 50, 60, 70, 80, 90, 100, 120, 140, 160, 180, 200.
Note that this set of points is different from the
set of points used in Fig.\ref{fig:rhor-points-P-1x8}, i.e.,
in this figure there is no points corresponding to 6, 7, 8, and 9 particles.
Note that in the lower temperature glass states, i.e., at $T \leq 0.004$
there develops an additional small peak at $(r/a) \approx 2.85$.
The inset shows the distributions of the coordination number of the particles.
Here the coordination number is defined through the position
of the minimum after the pre-peak of the first peak.
Thus, most of the particles have coordination numbers 4 or 5.
}\label{fig:rhor-points-P-4x0}
\end{center}
\end{figure}

Further, we discuss in more details the results obtained at pressures $P=1.8$ and $P=4.0$.
At these pressures, as follows from Fig. \ref{fig:expansion-coeff-1} and its discussion, 
we did not observe crystallization of the liquid, despite a rather 
long process of cooling implemented in our simulations.
Moreover, at these pressures we performed several increasingly
slow cooling runs and we still did not observe crystallization.
Note again that according to Fig. \ref{fig:gibbs0}, the pressures $P=1.8$ and $P=4.0$ are 
close to those pressures at which the transition from one crystal structure 
into a different one should happen as pressure increases at low temperatures.

The evolutions with temperature of the scaled PDFs,
$G_a(r/a)$ from (\ref{eq:pairdensityfunction}), 
at $P=1.8$ and $P=4.0$  are presented 
in Fig. \ref{fig:rhor-points-P-1x8},\ref{fig:rhor-points-P-4x0}.

The development of these PDFs, as temperature decreases,
exhibits a feature which is not usually observed in liquids 
with strong repulsion at short distances.
This feature is the splitting of the first peak into two with the decrease in temperature.
Thus, the single peak, observed at high temperatures, splits
into a pre-peak and the main peak as the temperature decreases.
We have already discussed a similar behavior in the context of
Fig. \ref{fig:pdf-select-01},\ref{fig:spdf2d0x009-1}, where
the splitting appears at high temperatures as the pressure increases.
Here, we discuss the splitting that happens at lower pressures as
the temperature decreases.
In order to demonstrate how the number of particles inside a sphere, 
centered on an average particle, increases with the increase 
of the radius of the sphere, we placed on the PDF curves
the solid circles which mark the specific values for 
the number of particles inside the average sphere.

In particular, as follows from Fig. \ref{fig:rhor-points-P-1x8}, at $P=1.8$, 
the inclusion of the first pre-peak, at $(r/a) \approx 1.0$,
corresponds to the inclusion of 6 or 7 particles,
while the inclusion of the main part of the first peak, at $(r/a) \approx 1.37$, 
adds another $\approx 12$ particles.
At $T=0.005$ and $T=0.004$ the splitting of the first peak
into the pre-peak and the main peak is essentially absent.
However, the splitting is observable at $T = 0.003$ 
and lower temperatures, i.e.,
there is no splitting in the liquid states, 
but there is a splitting in the glass states.

From  a qualitative perspective, the results at $P=4.0$, presented
in Fig.\ref{fig:rhor-points-P-4x0}, are similar to those at $P=1.80$.
However, from a quantitative perspective,
in comparison with Fig.\ref{fig:rhor-points-P-1x8},
the splitting of the first peak is already observable in the high-temperature liquid at $T=0.010$ and it
becomes very pronounced in the low-temperature glass states.
Further, at $P=4.0$, the number of particles which are inside 
the average sphere with the radius corresponding to the first 
minimum after the pre-peak is $\approx 4$, which implies 
a certain similarity to the diamond-like structure, 
in agreement with Fig. \ref{fig:gibbs0} and the results in Ref. \cite{Levashov20161}.
The inclusion of the main part of the first
peak (at $(r/a) \approx 1.48$) adds another $\approx 20$ particles.

A feature of particular interest, in our view, is the development of an additional peak, 
after the second peak, in the low-temperature glassy states.
At $P=1.80$, this peak develops at $(r/a) \approx 2.7$, while, at $P=4.0$, 
this peak develops at $(r/a) \approx 2.9$.
The splitting of the second PDF peak is a rather universal feature of 
supercooled liquids and glasses with strong interatomic repulsion \cite{Xu2011,Mizuguchi2009,Klumov2018}.
Here, we observe similar behaviour in the system with soft repulsion. 
Thus, it appears that the splitting of the second PDF peak it is a rather universal feature associated with the glass fromation.
Note that the splittings of the first and second peaks, observed 
in the HRPP liquid, are not related phenomena. 
While the former is caused by special features of the interatomic 
potential (ultrasoftness in our case), the latter appears to be a
general property, indicating that supercooled liquids and glasses 
demonstrate certain medium-range ordering on the length 
scales which involve (at least) the second neighbors.

\begin{figure}
\begin{center}
\includegraphics[angle=0,width=3.2in]{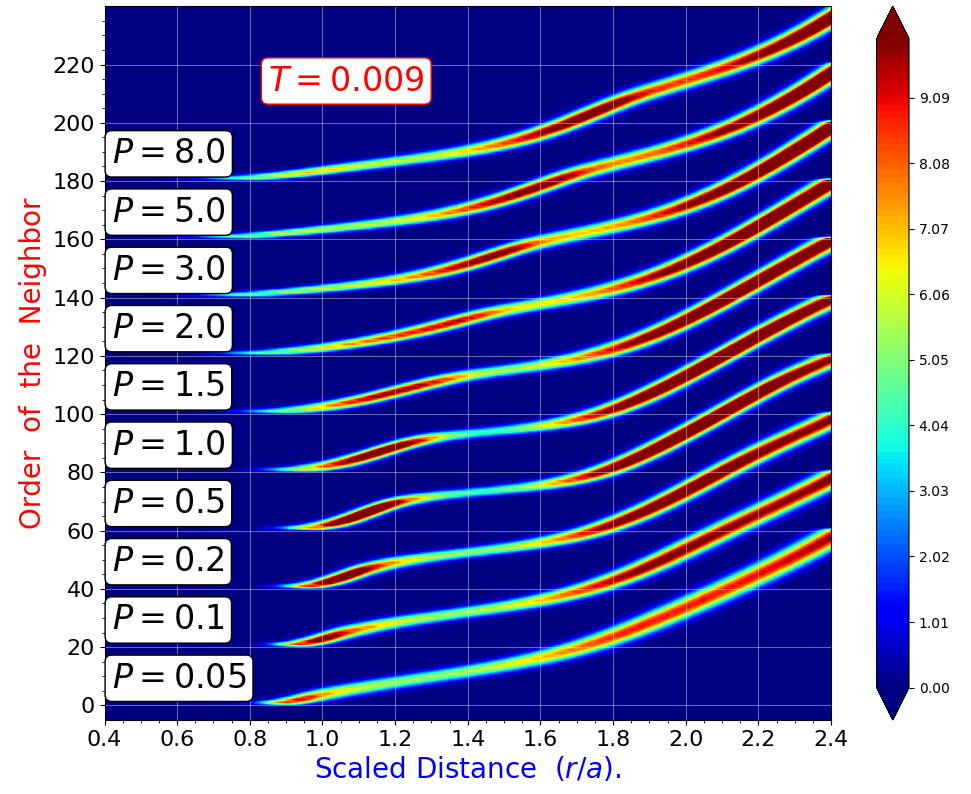}
\caption{The probability distributions for the pair distances at which the $n$-th
order neighbor for an average particle occurs.
Each curve for the next higher pressure is shifted upward (by 20) with 
respect to the curve at the previous lower pressure.
The shown range of the scaled distances, $(r/a)$,
approximately corresponds to the regions of the first and second peaks.
}\label{fig:order-neighbors-T-0x009}
\end{center}
\end{figure}
\begin{figure}
\begin{center}
\includegraphics[angle=0,width=3.2in]{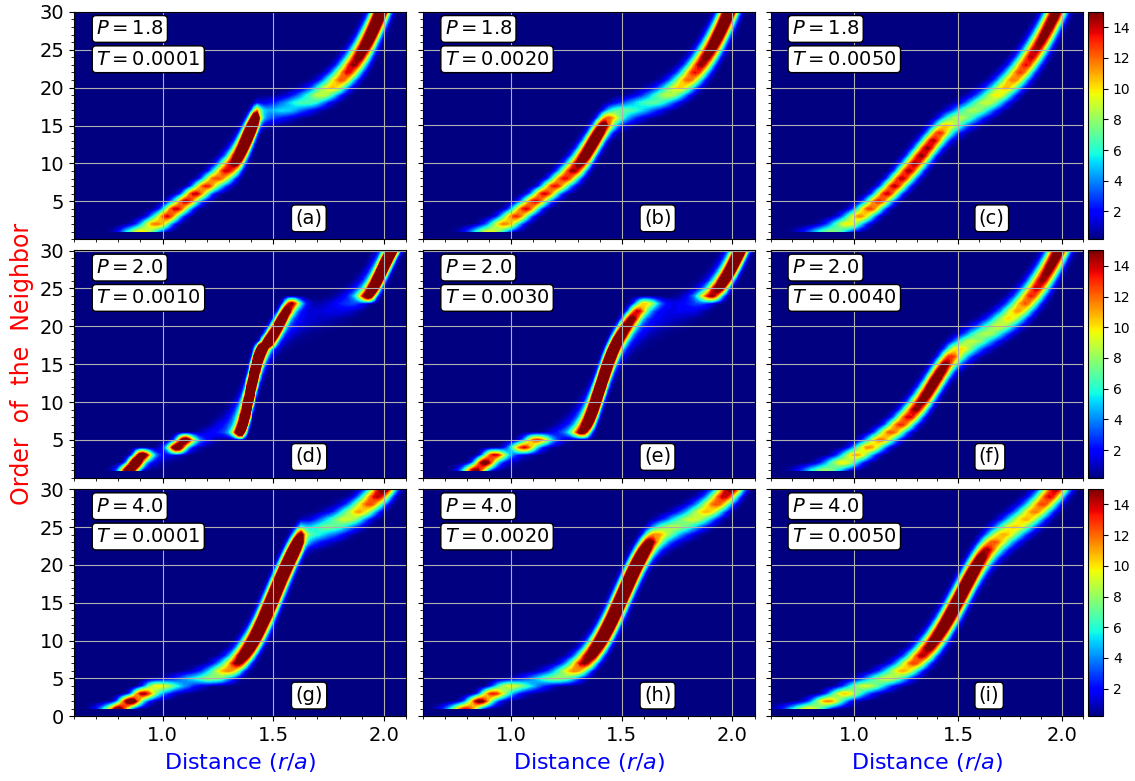}
\caption{The PDs for the pair distances at which the $n$-th order neighbor
for an average particle occurs at pressures $P=1.8$,  $P=2.0$, and  $P=4.0$ at selected temperatures.
At pressures $P=1.8$ and $P=4.0$ we did not observe crystallization of the liquid on cooling, i.e.,
the low temperature date correspond to the glassy state.
At pressures $P=2.0$ the liquid crystallized into the $Ia\bar{3}d$ lattice.
At $T=0.004$ ($P=2.0$) the system is in the liquid state, while at $T=0.003$ ($P=2.0$)
the system is in a ``high temperature" crystalline state.
At $T=0.001$ ($P=2.0$) the system is ``a low temperature" crystal.
}\label{fig:order-neighbors-P-2x0}
\end{center}
\end{figure}

Finally, we remind that while discussing the PDFs in section (\ref{sec:pair-distr}),
we anticipated that, at high pressures, the splitting of the second peak
should occur even at high temperatures.
While pressures $P=1.80$ and $P=4.00$ that we discuss here are lower than 
those at which we anticipated the splitting of the second peak 
in Fig. \ref{fig:spdf2d0x009-1}, 
it is clear that we discuss closely related structural behaviors.

It is well known that, in liquids composed of particles with strong short-distance repulsion, 
the first PDF peak corresponds to approximately 12-13 nearest neighbors \cite{Ryltsev2013PRE}; 
the number of particles with smaller (like 11) and larger (like 15) numbers of the nearest neighbors is rather small 
(see, for example, Fig. 2 from Ref. \cite{Levashov2008}).
In the HRPP liquid composed of soft particles the situation is quite different.
While this point can be demonstrated by simple integration of the pair density function, 
as it has been done for Fig. \ref{fig:rhor-points-P-1x8},\ref{fig:rhor-points-P-4x0}, 
we had considered also a somewhat different and more informative approach.
Thus, for every particle in the system, we ordered its neighbors, 
according to separation distance from the chosen particle.
Then, by averaging over the different chosen particles,
we created the probability distributions (PDs) to find $n$-th order
neighbor of the average particle at a certain distance.
These PDs are shown in Fig. \ref{fig:order-neighbors-T-0x009}.
It follows from the figure that, at high pressures, the number of 
neighbors associated with the first peak (which includes the pre-peak and the main peak) is noticeably larger than 12.
Thus, note that at $P=3.0$ and $P=5.0$ the number of the neighbors 
within distance $r < 1.6$ is approximately 20. See also Fig.\ref{fig:order-neighbors-P-2x0}.

The chosen way to represent ordering of the neighbors
is also useful to address how the structure changes with the temperature.
Thus, in Fig. \ref{fig:order-neighbors-P-2x0} we show the PDs to find $n$-th 
order neighbor at a given distance for the liquid (at $T=0.004$) and crystalline states at $P=2.0$.
At $P=2.0$, the system crystallizes into the $Ia\bar{3}d$ crystal 
with 3 first neighbors forming the equilateral triangle around the chosen particle, 
while 2 second neighbors are located above and below the triangle on one 
line with the chosen particle \cite{Levashov20161,Lokshin20181}.
The first and second neighbors can be clearly observed in panels (d,e).
We also see in these panels the small positive intensity in
the interval of distances $1.6 \leq (r/a) \leq 1.9$ and
in the interval of neighbors $21 \leq n \leq 24$, 
which shows the presence of particles whose environment deviates noticeably from the average structure.
Interestingly,  these deviations are associated with the particles
which effectively are the $3$-rd and the $4$-th neighbors in the $Ia\bar{3}$ lattice.

\subsection{Bond-Orientational order parameters}\label{sec:boo-analysis}

In this section, we address the structure of the HRPP system with the bond-orientational order 
parameters (BOOPs) \cite{Steinhardt19831}, which nowadays are routinely used \cite{Tanaka2018,Xu20141,Bagchi20181,Levashov20181,Klumov2016,Ryltsev2016}.
These parameters allow addressing the structural heterogeneity as they can be defined 
for each particle individually or a group of particles.
Usually, the BOOPs are used to describe the nearest neighbor environment of the individual particles.
In order to address the structure of the considered system at the selected conditions, 
we also use these parameters at the scale of the individual particles.
Since the BOOPs are well known, we omit their definitions, which can be found 
in the original paper of Steinhardt et al. \cite{Steinhardt19831}. 
By using these definitions, we almost completely reproduced
the results obtained for the several simple lattices \cite{boo-diff}. 
For the convenience of the reader, we provide in table \ref{table:BooQW} 
some reference values of the BOOPs, which will be useful for further analysis.

We start our discussion of the behavior of the BOOP in HRPP system from the considerations of the results
at $P=0.40$ and $P=1.40$.
The results at these pressures can be considered as the test cases 
that provide an initial insight into the behavior of the system.

\begin{figure}[t]
\begin{center}
\includegraphics[angle=0,width=3.5in]{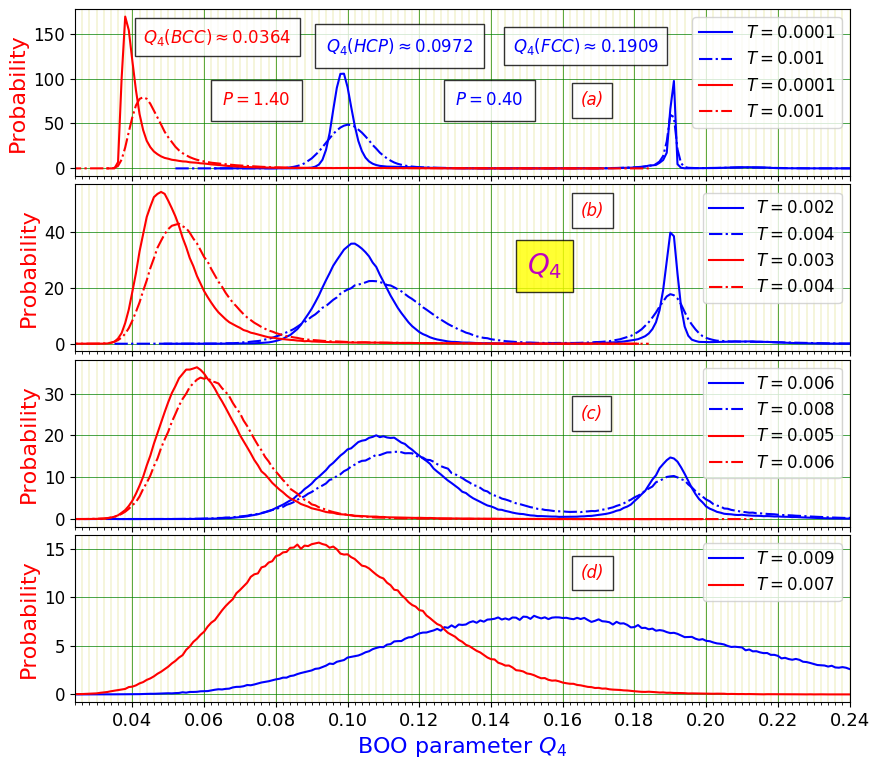}
\caption{
The probability distributions (PDs) of the BOO parameter $Q_4$ calculated for the particles at pressures $P=0.40$ and $P=1.40$.
At $P=0.40$ and at low temperatures ($T \leq 0.008$) the system is in a crystalline
state which is a mixture of the HCP ($\approx 75\%$) and FCC ($\approx 25\%$) lattices.
In particular, the thick layers of HCP are separated by the thinner layers of FCC.
The presence of the two lattices at $P=0.40$ is revealed by the two blue peaks in the PDs.
The positions of these peaks, as follows from table \ref{table:BooQW},
correspond to the HCP and FCC lattices.
At $P=1.40$ and at temperatures $T \leq 0.006$ the system is in the BCC state,
as follows from table \ref{table:BooQW} and the previous considerations.
Note that the PDs of the liquids above crystallization point are wide and featureless.
However, note also that the PD for the liquid at $P=0.40$ is clearly
on the right with respect to the PD for the liquid at $P=1.40$.
Also, note that the PDs for the HCP and FCC lattices are also on the right
with respect to the PD for the BCC lattice.
Thus, the liquid states up to some degree reflect the structures
into which these liquids crystallize on further supercooling.}
\label{fig:distr-Q4-P-0x40-P-1x40}
\end{center}
\end{figure}

The probability distributions (PDs) of the BOO parameter $Q_4$ at
pressures $P=0.40$ and $P=1.40$ are presented 
in Fig. \ref{fig:distr-Q4-P-0x40-P-1x40} for the selected temperatures.
The crystal lattices at low temperatures have been obtained
through the crystallization of the liquid states with 
further cooling at the corresponding constant pressure.

As it follows from Fig. \ref{fig:distr-Q4-P-0x40-P-1x40} 
and table \ref{table:BooQW}, at $P=0.40$ and 
at temperatures $T \leq 0.008$ [panels (a,b,c)] the system is in a ``crystalline" state 
consisting of a mixture of the HCP and FCC lattices with the dominant number of the HCP-type nearest neighbor environments.
The bond-orientational order analysis with the program OVITO (see Ref. \cite{ovito1,ovito2}) confirms this finding--it shows 
that $\approx 70\%$ of the particles have the local environment of the HCP lattice, 
while the rest have the local environment of the FCC lattice.
According to these results, the particles that have the HCP-like
environment form thick layers separated by thin layers of 
the particles that have the FCC-like environment.
In the liquid state [panel (d)] it is essentially impossible 
to assign to the particles any definite crystal-like environment due to a rather broad character of the distribution.
We remind here that at $T=0.009$ 
the system is rather close to the crystallization temperature.

\begin{center}
\begin{table}
\begin{tabular}{| c | c | c | c | c | c | c | } \hline
Lattice  & $Q_4$     & $W_4$       &   $Q_6$  & $W_6$  & $Q_8$ & $W_8$     \\ \hline
$SC$     & $0.7638$   & $0.1593$    & $0.5107$  & $0.0132$  & $0.7181$  & $0.0585$  \\ \hline
$FCC$    & $0.1909$   & $-0.1593$   & $0.5745$  & $-0.0132$ & $0.4039$  & $0.0585$  \\ \hline
$HCP$    & $0.0972$   & $0.1341$    & $0.4848$  & $-0.0124$ & $0.3170$  & $0.0513$  \\ \hline
$BCC$    & $0.0364$   & $0.1593$    & $0.5107$  & $0.0132$  & $0.4232$  & $0.0585$  \\ \hline
$ICOS$   & $0.0000$   & $---$    & $0.6633$  & $-0.1698$ & $0.0000$  & $---$ \\ \hline
\end{tabular}
\caption{The reference values of the BOO parameters for the selected lattices.
The values of the parameters $Q_l$ enter into the denominators of the definitions of the $W_l$.
Thus, since $Q_4$ and $Q_8$ for the icosahedral clusters (ICOS) are zero, 
the values of $W_4$ and $W_8$ for the icosahedral clusters are not defined.
}
\label{table:BooQW}
\end{table}
\end{center}

At $P=1.40$, as follows form Fig. \ref{fig:distr-Q4-P-0x40-P-1x40}(a,b,c) 
and table \ref{table:BooQW} the system is in the BCC state (or BCC-like state).
Note that the value of $Q_4$ corresponding to the BCC lattice is on
the left edge of the discussed probability distributions (PDs).
Note also that the PD for the BCC lattice is on the left
relative the PDs for the HCP and FCC lattices.
In the liquid state, at $P=1.40$ and $T=0.007$, 
it is difficult to make definite conclusions about the structure 
of the liquid, as the PD for $Q_4$ is rather broad.
However, note that the PD for the liquid at $P=1.40$ and $T=0.007$ 
is clearly on the left with respect to the PD for the liquid at $P=0.40$ and $T=0.009$.
Thus, there is a certain correlation between the structure
of the liquid state and the structure of the crystalline state 
into which the liquid crystallizes, as the PDs 
for the BCC lattice are also on the left relative the PDs for the FCC and HCP lattices.

\begin{figure}[t]
\begin{center}
\includegraphics[angle=0,width=3.5in]{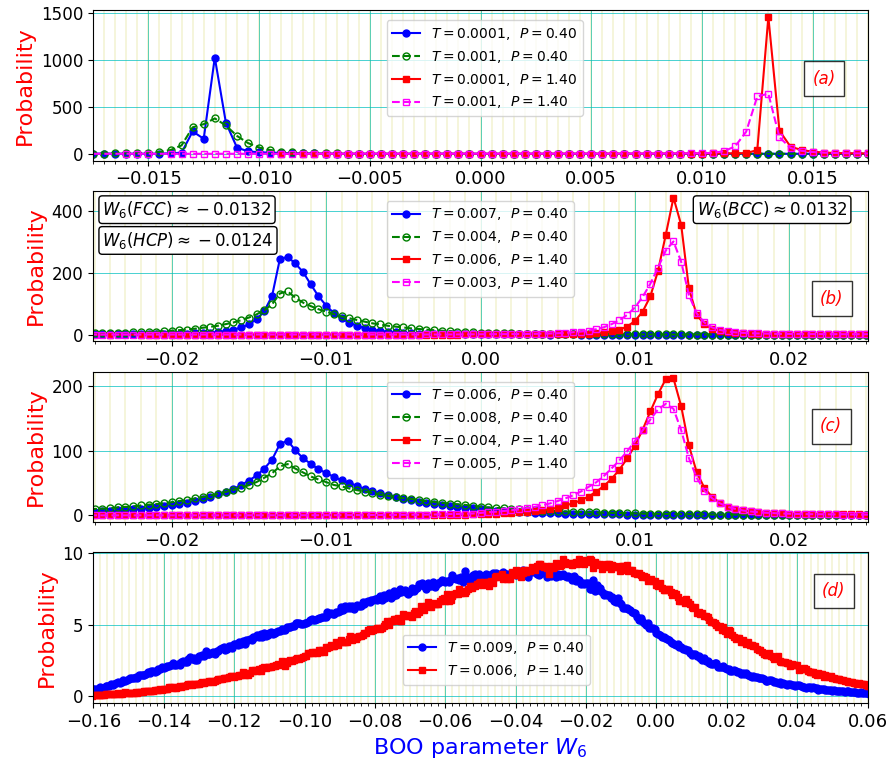}
\caption{
The probability distributions (PDs) of the BOO parameter $W_6$ calculated
for the particles in the same states as those discussed in Fig.\ref{fig:distr-Q4-P-0x40-P-1x40}.
According to table \ref{table:BooQW}, the values of $W_6$ for
the HCP and FCC lattices are rather close to each other: $W_6(HCP) \approx -0.0124$ and $W_6(FCC) \approx -0.0132$.
The small difference between these values can be observed
in the blue curve in panel (a), i.e., at the lowest studied temperature of $T=0.0001$.
At higher temperatures the parameter $W_6$ is not
capable to distinguish between the HCP and FCC lattices.
Note, however, that the parameter $Q_4$ in
Fig.\ref{fig:distr-Q4-P-0x40-P-1x40} clearly distinguishes between the two lattices.
Also, note that as the crystals melt the difference between
the PDs of $W_6$ is less pronounced than the differences
between the PDs of $Q_4$ in Fig.\ref{fig:distr-Q4-P-0x40-P-1x40}.}
\label{fig:distr-W6-P-0x40-P-1x40}
\end{center}
\end{figure}

The PDs of the BOO parameter $W_6$ are shown
in Fig. \ref{fig:distr-W6-P-0x40-P-1x40}.
Here, we discuss the behavior of $W_6$ parameter,
instead of the behavior of $W_4$, because for the BCC lattice,
even at low temperatures, the $W_4$ parameter has a rather wide and non-informative distribution.
According to table \ref{table:BooQW}, the values of $W_6$ for the FCC and HCP lattices are close to each other.
For this reason, 
the presence of two lattices can be guessed from the $W_6$ 
PD-curves only at the lowest temperature of $T=0.0001$,
i.e., in Fig. \ref{fig:distr-W6-P-0x40-P-1x40}(a)
there is a splitting of the blue peak into two (at $W_6 \approx -0.0128$).
At higher temperatures, the presence of the two lattices can not be
guessed from the behavior of the PD of $W_6$ alone.
The behavior of $W_6$ for the BCC lattice at $P=1.40$
exhibits regular behavior also for the crystal at the higher temperatures.
Panel (d) shows the behavior of $W_6$  for the liquids at $P=0.40$ and $P=1.40$ 
close to the respective crystallization points.
While the distributions for $P=0.40$ and $P=1.40$
are different, it is clear that these differences are not as
significant as the differences between
the probability distributions of $Q_4$ in Fig. \ref{fig:distr-Q4-P-0x40-P-1x40}(d).

\begin{figure}[t]
\begin{center}
\includegraphics[angle=0,width=3.5in]{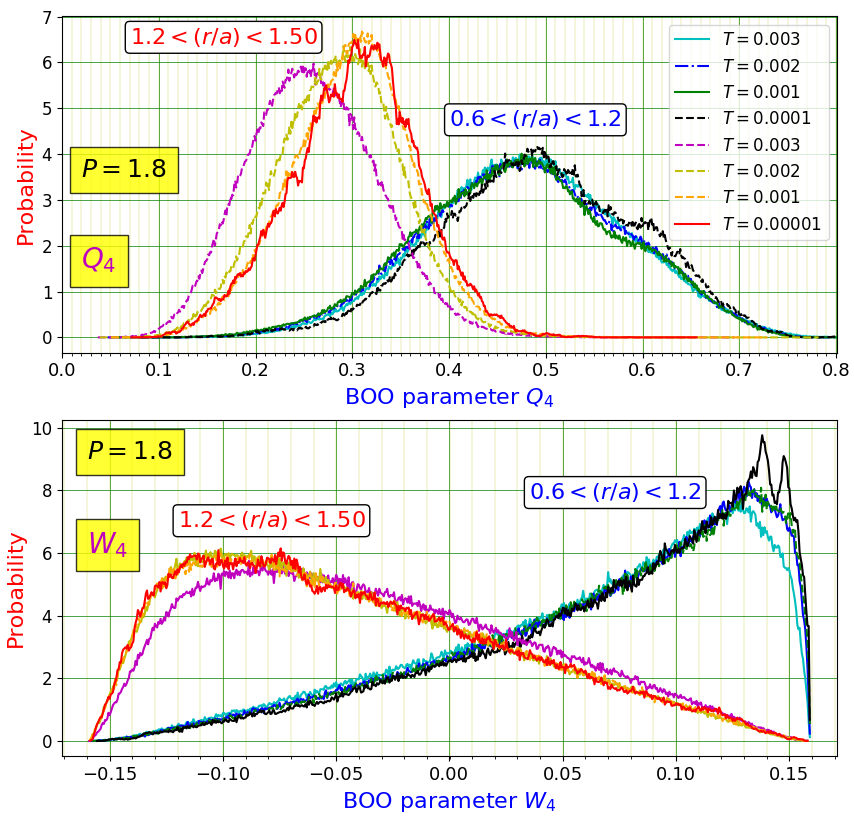}
\caption{The BOO parameters $Q_4$ and $W_4$ for the glass states obtained by cooling the liquid at $P=1.8$.
The choice of the glass states is dictated by the necessity to have a structure in which there is
an observable separation of the first peak into the pre-peak and
its main part--this separation is observable
in Fig. \ref{fig:rhor-points-P-1x8} for $T \leq 0.003$.
The results for the two distance regions are presented.
The first region, $0.6 \leq (r/a) \leq 1.2$, covers the position of the pre-peak of the first peak,
while the second region, $1.2 \leq (r/a) \leq 1.5$, covers the position of the main part of the first peak.
}\label{fig:boo-P-1x8-Q4-W4}
\end{center}
\end{figure}

\begin{figure}[t]
\begin{center}
\includegraphics[angle=0,width=3.5in]{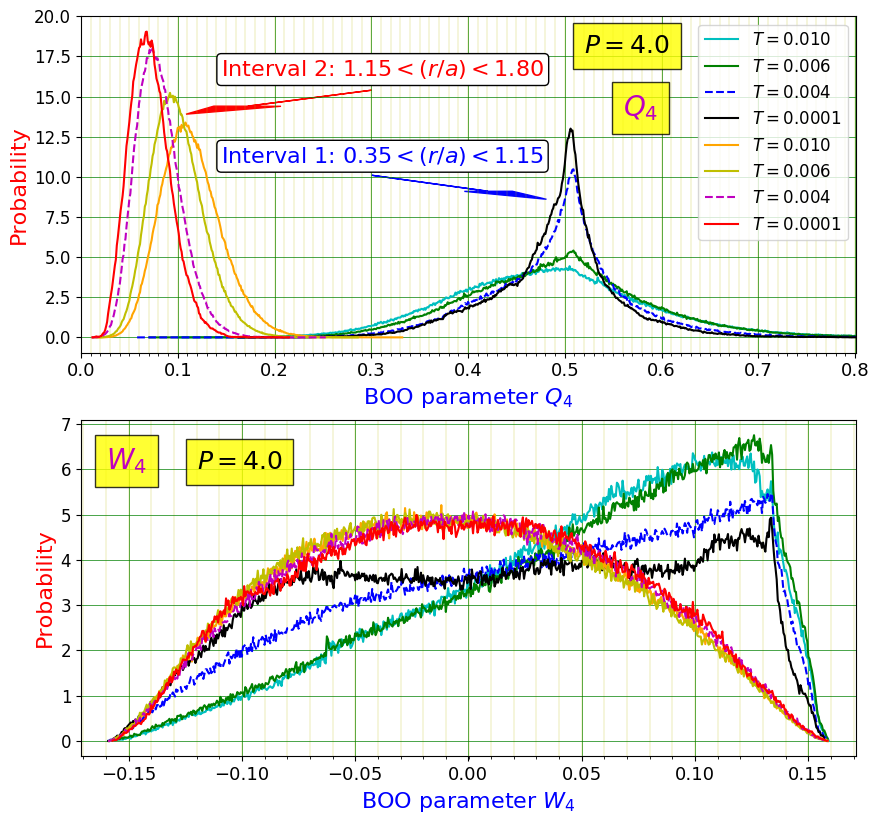}
\caption{
The BOO parameters $Q_4$ and $W_4$ for the liquid and glassy states
obtained by cooling the liquid at $P=4.0$.
The results for the two distance regions are presented.
The first region, $0.35 \leq (r/a) \leq 1.15$, covers the position of the pre-peak of the first peak,
while the second region, $1.15 \leq (r/a) \leq 1.8$,
covers the position of the main part of the first peak.
As follows from Fig.\ref{fig:rhor-points-P-4x0} and its inset, the number
of particles in the first region varies between
3 and 6 (mostly 4 and 5), while the number of particles associated with
the second region is $\approx 20$.
}\label{fig:boo-P-4x0-Q4-W4}
\end{center}
\end{figure}

Further, we discuss the results obtained at pressures of $P=1.8$ and $P=4.0$.
At these pressures, as follows from Fig. \ref{fig:expansion-coeff-1} and its discussion, 
we did not observe crystallization of the liquid, despite 
a rather long process of cooling implemented in the simulations.
Moreover, at these pressures we performed several increasingly
slow cooling runs and we still did not observe crystallization.
The PDs of the BOO parameters $Q_4$ and $W_4$ at pressures $P=1.80$ and $P=4.0$ 
are shown in Fig. \ref{fig:boo-P-1x8-Q4-W4},\ref{fig:boo-P-4x0-Q4-W4} 
for the selected temperatures.
In presenting these results, we consider separately
the contributions associated with the pre-peak of the first peak
and the contributions associated with the main part of the first peak.

At pressure $P=1.80$ the PDs of $Q_4$, associated with
the pre-peak [distance interval $0.6 \leq (r/a) \leq 1.2$],
exhibit very week temperature dependence. 
At the lowest temperature, one can observe the appearance of a shoulder at $Q_4 \approx 0.61$.
The center of the peak is located at $\approx 0.5$ and the comparison of this value with the values of $Q_4$ in table \ref{table:BooQW} leads to the conclusion that the bond-orientational structure associated with the pre-peak is closer to the simple cubic (SC) lattice than to any other lattice listed in the table (the PD is rather broad, of course).
This is in agreement with our previous discussion of Fig. \ref{fig:rhor-points-P-1x8}, i.e., with the fact that the region of the pre-peak contains approximately 6 particles.
Indeed, as follows from Fig.\ref{fig:rhor-points-P-1x8} and its inset,
at pressure $P=1.80$, the number of particles in the pre-peak region varies between 4 and 8 (mostly 5 and 7).
However, since the discussed PDs are rather broad, it is not reasonable to assume that the structure associated with the pre-peak is SC-like,
in our view.
The behaviors of the PDs of $Q_4$, for the main part of the first peak, exhibit more significant temperature dependence.
The comparison of the location of the PD peak ($Q_4 \approx 0.25$ or $Q_4 \approx 0.3$) with the values of $Q_4$ in table \ref{table:BooQW} shows that the BOO-structure of the main part of the first peak is more
FCC-like than any other of the presented lattices.
This is in agreement with the number of particles ($\approx 12$)
which can be associated with the main part of the first peak from Fig. \ref{fig:rhor-points-P-1x8}.
However, in our view, it is necessary to remember that the discussed PDs are rather broad.
The PDs associated with the $W_4$ parameter are also rather broad.
Note also that the locations of the maximums of the PDs of $W_4$,
after comparison with table \ref{table:BooQW},
support our interpretation that the BOO-structure of
the particles associated with the pre-peak of the first peak
resembles the SC structure, while the BOO structure of
the main part of the first peak resembles the FCC structure.


Figure \ref{fig:boo-P-4x0-Q4-W4} shows the PDs of the BOO 
parameters $Q_4$ and $W_4$, at $P=4.0$, for the selected temperatures.
In this case, the PDs associated with the pre-peak of the first peak 
exhibit a clear change as the liquid transforms into a glass.
Thus, as the temperature is reduced from $T=0.006$ to $T=0.004$,
there develops a BOO associated with the particles located 
in the pre-peak region.
Since the number of particles associated with the pre-peak 
is $\approx 4$, it is natural to associate this change with 
an abrupt enhancement of the tetrahedral order. 
Thus, at $P=4.0$, might be possible to associate 
the glass transition with an abrupt enhancement of the BOO.
The behavior of the peak associated with the main 
part of the first peak also exhibits a temperature dependence.
In particular, the distribution of the parameter $Q_4$, 
associated with the major part of the first peak, 
shifts to the region of smaller values, which means 
that the distribution of particles in this region becomes more orientationally
homogeneous.
Since the number of particles associated with the main part of 
the first peak is $\approx 20$, it might be reasonable to assume 
that these particles exhibit a tendency for ordering into the regular dodecahedron.
It is of interest to compare the results for the main part of the first peak in this figure and in Fig.\ref{fig:boo-P-1x8-Q4-W4}.
Thus, note that the distributions for $Q_4$, associated with the main peak in Fig.\ref{fig:boo-P-1x8-Q4-W4}, are shifted to larger values,
in comparison to the distributions presented in this figure. 
It is, indeed, natural to expect that more ordering should
be associated with $\approx 12$ particles than with  $\approx 20$ particles.

Concerning the probability distributions of $W_4$ for the pre-peak, 
we note that the PDs of $W_4$ exhibit a somewhat regular behavior 
at high temperatures, while in the low-temperature glass the PDs of $W_4$ are 
somewhat irregular and very broad.
It is difficult to interpret this behavior without more detailed considerations of the structures.
There is also some qualitative similarity with the results presented
in Fig.\ref{fig:boo-P-1x8-Q4-W4} for the pre-peak of the first peak.

\begin{figure}[t]
\begin{center}
\includegraphics[angle=0,width=3.5in]{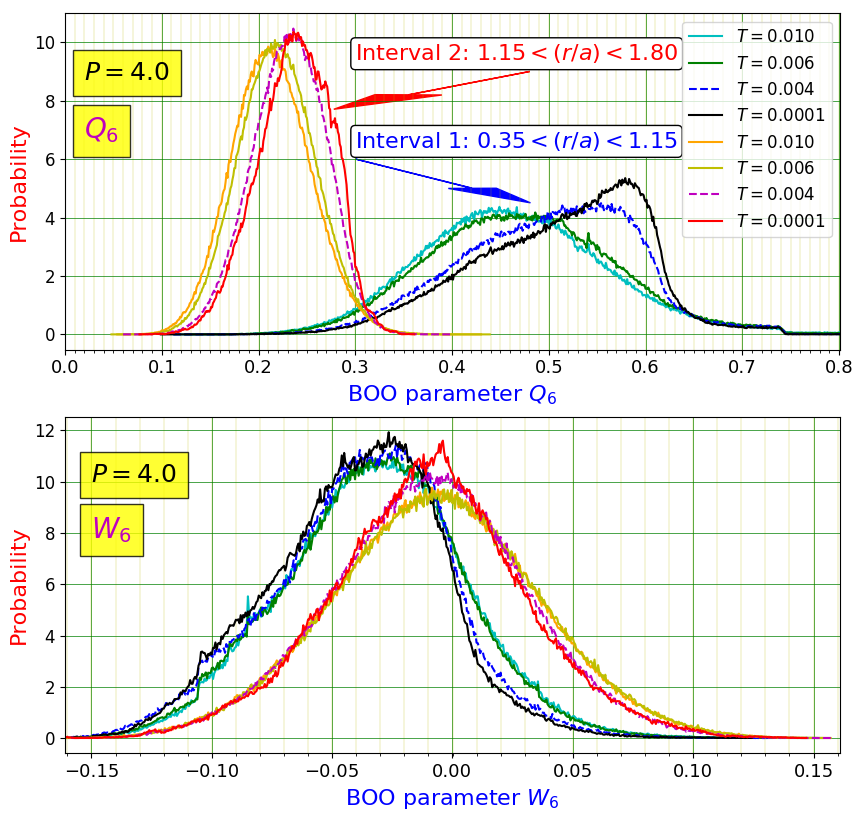}
\caption{
The BOO parameters $Q_6$ and $W_6$ for the liquid and glass states at $P=4.0$, i.e.,
for the same conditions and distance intervals at which have been calculated the data presented in Fig.\ref{fig:boo-P-4x0-Q4-W4}.
The largest differences between the low-temperature and high-temperature PDs
is observable for the parameter $Q_4$ and for the distance interval of the pre-peak of the first peak.
However, it is difficult to interpret these changes in a definite quantitative way.
The PDs of $W_6$ essentially do not change as the temperature is reduced and
thus the $W_6$ parameter is not informative in this case.
}\label{fig:boo-P-4x0-Q6-W6}
\end{center}
\end{figure}

Concerning the distributions of $W_4$ associated with the main part of the first peak we note that, 
in comparison with Fig.\ref{fig:boo-P-1x8-Q4-W4}, 
the character of the distributions of $W_4$ at $P=4.0$, 
i.e., in this figure, appear to be featureless meaning the absence of the structural ordering.
This is in agreement with the conclusions based on considerations of $Q_4$ distributions.

Finally, we note that the comparison of the PDs of the BOO parameters 
in Fig. \ref{fig:boo-P-1x8-Q4-W4} and Fig. \ref{fig:boo-P-4x0-Q4-W4}
shows that the observed PDs are quite different and thus the comparison shows
that the BOO parameters can capture the related differences.
However, due to the wideness of the PDs, it is not clear how to estimate the strengths
of the observed orderings without more detailed considerations and the modeling of the structures.

Figure \ref{fig:boo-P-4x0-Q6-W6} shows the PDs of the BOO parameters $Q_6$ and $W_6$ at $P=4.0$ at the selected temperatures.
The most pronounced changes with temperature happen with the PDs of $Q_6$ for the pre-peak of the first peak.
The shape of the PD at the lowest temperature may suggest the existence of the two
possible arrangements of the nearest 4 particles. However, it is impossible 
to draw any definite conclusions without a more detailed modeling. 
This modeling is indeed necessary for the estimation of the strengths of the observed orderings.

Based on the conducted investigations with the BOO parameters,
we conclude that it is possible to draw certain conclusions 
from the behavior of these parameters.
However, to estimate the strengths of the observed BOO correlations,
it is necessary to make comparisons to some reference cases and
thus it is necessary to conduct more detailed modeling,
which we consider being beyond the scope of this paper.

\section{Conclusion}\label{sec:conclusion}

In this paper, we studied how the structure of the single-component liquid composed of particles interacting through
the harmonic-repulsive pair potential depends on the pressure in a relatively wide interval of pressures
and at temperatures which are close to the simulated crystallization temperatures.
In our previous study, we demonstrated that this simple liquid,
at different densities, crystallizes into several different crystal structures 
(some of them are quite complex).
Thus, it is reasonable to expect that the structure of the liquid, 
as pressure changes, also can exhibit significant variations.

Two methods have been used to address the liquid's structure.
In particular: 
a method based on considerations of the properly scaled pair density function
and a method based on consideration of the bond-orientational order parameters.
The emphasis of our studies was on the structural
changes which go beyond the simple rescaling of the inter-particle distances.

At first, we studied the liquid from a macroscopic perspective and observed, at some pressures, several unusual properties.
In particular, we found that, at certain pressures, the liquid demonstrates water-like anomalies, i.e., negative
temperature expansion coefficient and an increase of the volume upon crystallization. We also found that at
some pressures the potential energy of the system increases as the temperature of the system decreases.

The single-component systems with strong repulsion between the particles at short distances 
usually easily crystallize as the liquid enters into the supercooled regime.
This is not the case for the studied system at all pressures.
Thus, as can be expected from the previous study, we found that 
at certain pressures the system exhibits remarkable stability against crystallization.
Therefore, particular attention has been devoted studies of the liquid at these pressures.
It was demonstrated that the pressure regions of unusual behavior and 
the stability of the liquid against crystallization occur 
at the boundaries between the different crystal structures that can be produced from the liquid on cooling.

Our studies with the scaled pair density function (SPDF) show that the most significant structural change,
that happens with the liquid as the pressure increases, occurs at a density at which there appears 
non-negligible interaction with the second neighbors.
As this happens, the first peak of the scaled PDF splits into the pre-peak and the main peak. 
This is also a characteristic feature of the previously observed crystal structures, for example, the $Ia\bar{3}d$ structure.
Therefore, non-negligible interaction with the second neighbors should cause the appearance of
complex crystal structures in the crystallization process and also lead to the remarkable stability of the liquid against crystallization.
Indeed, it is reasonable to assume that, as there appears interaction with the second neighbors,
that there also appears competition between the more complex ground states.
The competition between these complex structures effectively also 
should create frustration that prevents crystallization.

In our investigations with the bond-orientational order (BOO) parameters, we, at first,
studied the behavior of these parameters in two reference cases, i.e., for the pressures at which
the HCP, FCC, and BCC lattices form from the liquid state on cooling.
These considerations allowed us to estimate the characteristic wideness
of the peaks in the probability distributions (PDs) of the BOO parameters.
Then we considered the liquid and the glass states at pressures at which the liquid exhibits
remarkable stability against crystallization.
We found that the PDs of the considered BOO parameters are rather wide, and thus
they do not show the formation of some well expressed orientational order.
A possible exception from this general observation is related to
the formation of the tetrahedral arrangements around particles at $P=4.0$,
as can be seen the most clearly from the behavior of the $Q_4$ parameter at this pressure on cooling.

Since the probability distributions of the BOO parameters often are rather wide, 
the obtained data suggest that other methods, for example, 
the bond-orientational order diagrams (BOOD) might be useful 
in addressing the structures of ultrasoft liquids \cite{Glotzer20181},\cite{Engel20091}.

The most important conclusion we make from the obtained results is that single-component systems with
one-scale ultrasoft potentials also can demonstrate non-trivial behavior usually observed
in more complex multiscale systems with core-softened or oscillating potentials.
In the end, we suggest that the presence of non-negligible interactions with
the second and further neighbors, causes such a complex behavior in all of the mentioned systems.


\section{Acknowledgements}

This work was supported by Russian Science Foundation (grant RNF 18-12-00438).



\begin{thebibliography}{10}
\bibitem{boo-diff}
In our calculations we also observed
certain differences with the results presented therein.
However, these differences appear also to coincide with some inconsistencies
in the original publication and thus we believe that our results are correct.
For example, the value of the parameter $W_6$ for the FCC lattice 
in Fig.2 of Ref. \cite{Steinhardt19831} is inconsistent
with its value in table 1 of the same reference. 
Then, there is a clear contradiction between the value of $W_6$ for the
BCC lattice in the just mentioned figure and the table. Another issue that we noticed is related to the value of $W_4$ of the BCC lattice.
It is possible to consider separately the values of $W_4$ in the BCC lattice for the very first sub-shell, consisting of 8 particles,
and for the second sub-shell, consisting of 6 particles. 
As it turns out, the absolute values of $W_4$ for these two sub-shells
are the same, but their signs are opposite.
For this reason, the distribution of the values of $W_4$ for the full first 
shell (in total 14 particles) of the BCC lattice turns out to be rather broad, 
as various deformations of the first and second 
sub-shells can lead to the broad spectrum of the results.
\end{thebibliography}
\end{document}